 \documentstyle[11pt,aaspp4,tighten]{article}

\received{1996}               
\accepted{1996}
\journalid{date }{}       

\slugcomment{draft of \today}

\def\etal{{\it et al.~}}
\def\eg{{\it e.g.~}}
\def\ie{{\it i.e.~}}

\def\cm3{~{\rm cm^{-3}}}

\def\Msun{~{\rm M}_{\sun}}
\def\lsim{\raise0.3ex\hbox{$<$}\kern-0.75em{\lower0.65ex\hbox{$\sim$}}} 
\def\gsim{\raise0.3ex\hbox{$>$}\kern-
0.75em{\lower0.65ex\hbox{$\sim$}}} 
\def\ha{H$\alpha$}

\def\oi{[O\thinspace {\scriptsize I}]}
\def\sii{[S\thinspace {\scriptsize II}]}

\begin{document}

\font\twelvei = cmmi10 scaled\magstep1 
       \font\teni = cmmi10 \font\seveni = cmmi7
\font\mbf = cmmib10 scaled\magstep1
       \font\mbfs = cmmib10 \font\mbfss = cmmib10 scaled 833
\font\msybf = cmbsy10 scaled\magstep1
       \font\msybfs = cmbsy10 \font\msybfss = cmbsy10 scaled 833
\textfont1 = \twelvei
       \scriptfont1 = \twelvei \scriptscriptfont1 = \teni
       \def\mit{\fam1 }
\textfont9 = \mbf
       \scriptfont9 = \mbfs \scriptscriptfont9 = \mbfss
       \def\bmit{\fam9 }
\textfont10 = \msybf
       \scriptfont10 = \msybfs \scriptscriptfont10 = \msybfss
       \def\bmsy{\fam10 }

\title{$10^{51}$ Ergs: The Evolution of Shell Supernova Remnants}
                    
\author{T. W. Jones }                            
\author{Lawrence Rudnick}
\and
\author{Byung-Il Jun}
\affil{Department of Astronomy, University of Minnesota,                
    Minneapolis, MN 55455}
\author{Kazimierz J. Borkowski}
\affil{Department of Physics, North Carolina State University, Raleigh,
	NC 27695-8202}
\author{Gloria Dubner}
\affil{Instituto de Astronomia y Fisica del Espacio. Buenos Aires, 
Argentina}
\author{Dale A. Frail}
\affil{National Radio Astronomy Observatory, Socorro, NM, 87801}
\author{Hyesung Kang}
\affil{Department of Astronomy, University of Washington,
    Seattle, WA 98195-1580}
\author{Namir E. Kassim}
\affil{Naval Research Laboratory, Washington, D. C. 20375-5351}
\and
\author{Richard McCray}
\affil{JILA, University of Colorado, Boulder, CO, 80309-0440}
    \bigskip
                    
\clearpage
                    
\begin{abstract}  
This paper reports on the workshop, {\it $10^{51}$ Ergs: The Evolution 
of 
Shell 
Supernova Remnants},
hosted by the University of Minnesota, March 23-26, 1997. The
workshop was designed to address fundamental dynamical issues associated
with the evolution of shell supernova remnants and to understand
better the relationships between supernova remnants and their environments. 
Although the title points only to classical, shell SNR structures, the 
workshop
also considered dynamical issues involving X-ray filled composite
remnants and pulsar driven shells, such as that in the Crab Nebula.
Approximately 75 observers, theorists and numerical simulators with
wide ranging interests attended the workshop. An even
larger community helped through extensive on-line
debates prior to the meeting to focus issues and galvanize discussion.

In order to deflect thinking away from traditional patterns,
the workshop was organized around chronological sessions for ``very 
young'',
``young'', ``mature'' and ``old'' remnants, with the implicit
recognition that these labels are often difficult to apply. 
Special sessions were devoted to related issues in plerions
and ``thermal X-ray composites''.
Controversy and debate were encouraged. Each session also
addressed some underlying, general physical themes: How
are SNR dynamics and structures modified by the character of the
CSM and the ISM and vice versa? How are magnetic fields generated
in SNRs and how do magnetic fields influence SNRs?   Where and how are
cosmic-rays (electrons and ions) produced in SNRs and how does their
presence influence or reveal SNR dynamics? How does SNR blast energy 
partition into various components over time and what controls
conversion between components?
In lieu of a proceedings volume, we present here a synopsis of the
workshop in the form of brief summaries of the workshop sessions.
The sharpest impressions from the workshop were the crucial and
under-appreciated roles
that environments have on SNR appearance and dynamics and the
critical need for broad-based studies to understand these
beautiful, but enigmatic objects.
\end{abstract}      
                    
                    
\section{Introduction}\label{sec:intro}

Supernova remnants (SNRs) play dramatic and essential roles in the 
dynamics of the interstellar medium. They probably supply
all but the highest energy cosmic-rays, and they are the means
by which heavy elements produced during supernovae are introduced to
the ISM. The SNR paradigm for the past quarter century has been based
on Woltjer's 1972 cartoon describing SNRs as spherical shells
in one of four distinct phases  of expansion into uniform media
These phases have come to be known as ``free expansion'', ``Sedov, 
adiabatic
blast wave'', ``radiative snow-plow'', and ``dispersal''. Yet, the 
evidence 
is strong that this cartoon is inadequate as a model for real SNR 
dynamics. The distinct phases imagined by Woltjer may be brief or 
not occur at all in a given remnant. Further, SNRs are generally not 
spherical and not
interacting with uniform media.
Thus, different dynamical stages may occur simultaneously within
a single remnant, and structures may be very confusing.
The observational and theoretical evidence of this dynamical 
complexity  seriously limits our ability to
decipher critical issues ranging from SNR age determination to their
role in establishing the structure of the ISM. 
This workshop was conceived to define key questions needing answers
for progress in these matters.

For convenience, and also to avoid conventional labels (\eg, ``Sedov''), 
most of the workshop was organized around chronological sessions for 
``very young'', ``young'', ``mature'' and ``old'' remnants, with the 
implicit
recognition that these labels are often difficult to apply. 
Each session addressed some underlying, general physical themes: How
are SNR dynamics and structures modified by the character of the
CSM and the ISM and vice versa? How are magnetic fields generated
in SNRs and how do magnetic fields influence SNRs?  Where and how are
cosmic-rays (electrons and ions) produced in SNRs and how does their
presence influence or reveal SNR dynamics? How does SNR blast energy
partition into various components over time and what controls
conversion between components? Several months in advance of the actual 
meeting a web site was established (http://ast1.spa.umn.edu/SNRmeet) 
where participants were encouraged to  raise questions that they wanted 
answered and to debate issues. This was very successful in ``setting the 
stage'', so that many preliminary matters had been resolved or focussed.  
The web site 
continues to be accessible for review of the pre-meeting discussions, as 
well as the submitted abstracts.

Although the simple, chronological organization was sometimes
awkward, heated controversy over this classification scheme never 
materialized. The obvious short-comings of a purely time-based 
categorization were raised in the web-based discussions and during 
introductory talks. This highlighted the complexity 
associated with finding unifying physical concepts.
In terms of the traditional dynamical cartoon, there was broad 
agreement that no discrete transitions exist between the different
evolutionary stages.  So, some additional simplifying theme would
be needed to organize remnants  according to 
dynamical age. In response to this, 
{\it Lozinskaya}\footnote{Throughout this paper, names in italics
refer to meeting contributions.  A key to contributors' names
and institutions is found in the Appendix.}
suggested the reduced radius, $R \times \rho^{1/3}$, 
as a better indicator of evolutionary age than
physical radius itself (see Lozinskaya 1992). 
Alternatively, the mass ratio (swept-up mass divided
by ejecta mass) as an age indicator was emphasized (Jun
and Norman 1996a; {\it Dohm-Palmer \& Jones; Jun} ). 
Interpretation of this 
mass ratio is also dependent on the
ejecta density distribution, as pointed out by {\it Dwarkadas}. 
Consequently, it is probably most  
meaningful to define the mass ratio as the swept-up mass divided by
the ejecta mass contained only in the shell.
What is more confusing in age determination is that even single remnants 
show
different evolutionary ages as a function of azimuthal angle
(\eg {\it Koralesky \& Rudnick; Shull}).  Another suggested tool 
applied frequently during the workshop was
the so-called expansion parameter, $m$ ($R \propto t^m$), which
measures shell deceleration, and, hence, the accumulated interaction
with ambient media.  However, that measure is also seen to vary strongly
within a given SNR.

	The pre-workshop on-line discussions also included
active debates about the nature of X-ray filled remnants, and 
interpretations of shell-like features in pulsar-driven remnants.
These issues seemed important enough and close enough to the original
intent for the workshop that sessions were then set aside for them.
During discussion {\it Lozinskaya} also raised the question of whether these plerions
represent an evolutionary stage of shell remnants.

On balance, the age classifications did prove quite useful in guiding
our discussions. For example, there
is no question that  radio light curves of SNe shown by {\it Montes} 
were
in fact ``Very Young'' SNRs, and nobody objected to referring to the
historical SNRs such as SN1006, Tycho, and Cas-A as being ``Young''.
Conversely it was logical to refer to the largest Galactic loop-like
structures discussed by {\it Heiles} as ``Old'' SNRs. This, of course, 
relegated
the majority of observed SNRs (as well as the majority of papers
presented at this meeting) to the category of ``Mature'' remnants.

Two physical issues so broadly transcended the age divisions that 
they
were given small sessions unto themselves. These were the relationship
between SNRs and the physics of particle acceleration (\ie, ``cosmic-
rays'')
and the generation of magnetic fields through instabilities in SNRs.
Both of these issues are critical to the interpretation of SNR 
observations,
while SNRs also seem to be uniquely useful laboratories for 
understanding
some very complex associated physics. 

At the workshop itself, there were reviews and talks highlighting new results,
 considerable discussion, and poster presentations.
Impromptu evening discussions were also held on specific themes of interest
to participants.  In this paper, we seek to summarize the broad
range of information and issues raised at the workshop.
The sessions are reviewed in \S 2-9, while \S 10 presents parting
messages from participants to the SNR community.

\section{Very Young Remnants} \label{sec:veryyoung}

One of the major recurring themes of the workshop was the critical
importance of environment to SNRs, their dynamics and appearance.
Thus the intent of a focus on ``very young'' 
remnants was to isolate those that are
still interacting entirely with their own circumstellar environments.
The significance of that CSM to the dynamics and appearance of very young
remnants was clearly emphasized during the workshop.
The discussion that took place can be divided
into SN 1987A and then all other very young remnants.  
The physical distinction between very young and young remnants was not
clearly resolved during discussions.
In this summary, we simply assume the age of very young
remnants to be in the range between $0 \leq t \leq 50$ years.

{\it Chevalier}, in a review,  pointed out
that most very young remnants are radio emitters.  This radio
emission can be explained by the power-law-density supernova
envelope interacting with circumstellar matter (Chevalier 1982b).
The interaction between the supernova ejecta and circumstellar matter
generates a double-shock structure that includes, from the outside in,  
a 
blast 
wave, contact
discontinuity, and a reverse shock.
The resulting shell is Rayleigh-Taylor (R-T) unstable and the
instability is believed to amplify the preexisting magnetic field.
The instability has been found to produce turbulent mixing during the
nonlinear instability stage (Chevalier, Blondin, and Emmering 1992) and amplify the
magnetic field to enhance the bright radio shell (Jun and Norman
1996a,b).  X-ray emission can also be explained by this model.
The \cite{che94} model explains X-ray emission from the remnant by
a reverse shock propagating into the ejecta.  
The reverse shock produces
higher temperatures for a flatter supernova density profile because 
a faster reverse shock front results.   Radiative cooling with high
enough densities to produce the optical emission can be
important near the reverse shock.
  The growth of the R-T instability in the
radiating shell is found to be higher than in the adiabatic case
(Chevalier and Blondin 1995).   Since the density profiles in the
ejecta and circumstellar medium determine the evolution and the
physical structure of the shell, the study of emission can yield 
information on the ejecta and the CSM density distribution.
It is also important to note that the distribution of
circumstellar matter can change shell morphologies
such as in the remnants of SN 1986J and 41.9 +58 in M82. These SNRs show 
a 
dramatic deviation from spherical symmetry (for possible mechanisms see 
Blondin \etal 1996; Jun \etal 1996).

\subsection{SN 1987A}\label{sec:1987A}

{\it Gaensler} presented an observational study of the radio emission 
from the remnant of SN 1987A.   After an early peak of the radio flux 
density, a 
second turn-on has been detected 1200 days after the explosion. That 
has been 
followed by a monotonically increasing flux density.   
The spectral index is found to be very steep and to remain constant at 
$\alpha = -0.95~(S_{\nu} \propto \nu^{\alpha})$.  This
spectral index is much steeper than predicted by first order Fermi
acceleration at a strong shock ($\alpha = -0.5$) (see \S \ref{sec:accel}).
Duffy \etal (1995) have modeled a cosmic-ray modified shock, using its 
reduced subshock velocity jump to explain this steeper spectrum. 
However, the reduced compression ratio of the cosmic-ray mediated shock 
may not  heat the gas enough to produce X-rays (as pointed out by {\it 
Jun}).
Another concern with this model is the need for a time-constant
modification to the shock, whereas spherical shocks steadily evolve.
(See \S \ref{sec:accel} for more on these kinds of issues.)
Expansion of the radio remnant in SNR 1987A has been measured to be
very slow, with the expansion factor $m = 0.11$, and the 
velocity $v = 2400$ km s$^{-1}$.  These imply a large
deceleration of the shock front.   It has been suggested that the
shock has encountered an HII region inside the optical ring (Chevalier 
and Dwarkadas 1995).  Assuming a constant rate of shock
in the HII region,
{\it Gaensler} predicts that the shock will hit the 
ring in the year $2008 \pm 3$ and 
that SNR 1987A will become a strong source in X-ray, optical, IR, UV, 
and radio.
{\it Gaensler} showed  the radio morphology already observed for
SNR 1987A. It contains
two ``hotspots'' to the east and west.  They are aligned along
the major axis of the optical ring.  He suggested that this morphology
is likely due to an axisymmetric circumstellar medium.
{\it Borkowski, Blondin and McCray} presented two-dimensional
hydrodynamic simulations of the interaction of the shock with the HII
region (Borkowski et al.~1997).  
The soft X-ray spectrum should be
dominated by emission lines of hydrogenic and helium-like
C,N,O, and Ne.  In addition, they predicted that broad  Ly$\alpha$ and
{N~V} $\lambda$ 1240 should brighten
steadily and should be observable with the Hubble Space Telescope during
the collision.
{\bf N.B.} Shortly after the workshop, the broad Ly$\alpha$ emission was
indeed detected with the HST (Sonneborn et al.~1997;
Michael et al.~1997). The impact of the
blast wave with the dense material at the inner edge of the ring was also
detected with the HST in the optical (Sonneborn et al.~1997; Garnavich,
Kirshner, \&\ Challis 1997).

{\it Drake} with {\it Glendenning, Estabrook, McCray, Remington,
Rubenchik, Liang, London, R. Wallace, and Kane} 
reported on progress in laboratory
experiments relevant to structure in SN 1987A.  
By using the Nova Laser to drive high-Mach-number ejecta into
ambient plasma, they investigated the formation of
a strong shock and its  reflected shock.
The experimental data show
reasonably good agreement with analogous one-dimensional numerical 
simulations.
Drake and his collaborators plan subsequently to study the nonlinear R-T
instability experimentally.  In an independent study,
also using the Nova Laser, Kane \etal 1997 have presented experimental 
investigation of the Richtmyer-Meshkov instability and Rayleigh-Taylor
instability that resemble the mixing in the He-H interface of SN 1987A.

\subsection{Other Very Young Remnants}\label{sec:ovyr}

{\it Montes} summarized the study of the radio emission from many 
different
very young supernova remnants.  By using Chevalier's ``mini-shell'' 
model
(Chevalier 1982b), he described how model fitting can determine
the physical properties of radio supernovae; namely, the  supernova
density profile ($\rho \propto r^{-n}$), density fluctuations in a  
circumstellar wind, and the presupernova mass loss rate.
Two well-studied examples are SN1979C and 1980K. The time variation of
radio surface brightness of SN1979C is found to be
oscillating with a period 1575 days.   Montes interpreted this as a
variation in the circumstellar density resulting from a modulation of
the presupernova mass loss rate.  Taking a shock velocity $ \sim 10^4 $
km s$^{-1}$
and a wind velocity $\sim 10$ km s$^{-1}$, the presupernova wind density
modulation appears to have had a time scale of $\sim 4000$ years.
He suggested a stellar companion to the presupernova star in a highly
eccentric orbit with period $\sim 4000 $ years as the most plausible 
mechanism for this wind modulation.  After 4300 days, SN1979C shows a
flattening of the radio light curve.  At about 6500 days, the radio
surface brightness is about 1.8 times higher than an
extrapolation from the earlier evolution.   
This corresponds to about 1.4 times higher density in the
circumstellar medium than expected.  On the other hand, SN1980K shows
a sharp drop in radio emission at about 4000 days.  The surface
brightness at about 6000 days is only about half the extrapolated value.
This implies about 0.6 times the expected density in circumstellar
medium.   The implied time scale for the sudden change in wind
structure revealed by radio surface
brightness of SNe 1979C and 1980K is comparable to the evolution time 
from
red supergiant to blue supergiant for $12-14~ \Msun$
stars. Montes 
concluded that radio monitoring can detect and characterize short
time-scale, presupernova stellar evolution.

{\it Van Dyk, Montes, Sramek, Weiler, and Panagia} presented a progress
report on the recent very young remnants, SNe 1993J and 1996cb (Type
IIb), SNe 1994I, 1996N, and 1997X (Type Ic), and SN 1995N (Type IIn).
They find that the model fit for SN 1993J implies $\rho_{csm} \propto
r^{-3/2} $ where $\rho_{csm}$ is the density of circumstellar gas.
They interpret this as a decrease in the pre-SN mass loss rate or
increase in wind velocity before explosion.  Also, the rate of
increase in the early radio emission requires the presence of higher
density ``clumps'' in the stellar wind (Van Dyk \etal 1994).  {\it 
Arnett}
showed two-dimensional numerical simulations of the interaction
between a red supergiant wind and a blue supergiant wind that
develops after the outer envelope of a massive star is shed. The 
resulting
medium is clumpy due to instabilities. He pointed out, therefore, that 
the assumption of a uniform mass distribution in the
circumstellar medium  during the expansion of a very young SNR blast
may not be at all a good approximation.

{\it Fesen} talked about the optical emission from various types of
very young supernova remnants.
Supernovae type II-Linear (SN 1970G, 1979C, 1980K, and 1986E)
show broad emission lines of H$\alpha$, [OI], [CaII], [FeII], and
[OIII].   These optical emissions come from the shocked and
photoionized SN ejecta according to the model of the supernova
interacting with circumstellar matter. However, some of the predicted
features are not observed (\eg a steep H$\alpha$ flux decline and
[OIII] domination in the spectrum at late times ).   Supernovae type II
pec (SN 1978K, 1986J, and 1988Z) show narrow emission lines and dominant
H$\alpha$ emission at late times.  These features can also be
modeled assuming SN-CSM interaction in which the CSM includes dense 
clouds ($\rho \sim 10^{6 - 7} cm^{-3}$).
In this model, the strong H$\alpha$ emission comes from the trailing
HII region downstream of the transmitted shock in the cloud.
Fesen questioned why fast ejecta cannot be seen at late times.  He also
puzzled over the missing optical emission from the very young-
young supernova remnant transition period that is typically 
about $50 \sim 300$ years after the SNe. 
This is an important problem in understanding the
connection between supernovae and supernova remnants.

\section{Young Remnants}\label{sec:young}

Young SNRs occupy a special place in studies of stellar explosions and 
their
influence on the ISM. These bright objects have been intensively studied
throughout the electromagnetic spectrum. In a number of cases the SN 
explosions themselves had been recorded in the past, and their remnants 
are 
still in a phase of a rapid dynamical evolution. Stellar ejecta in young 
SNRs 
should still be visible, which makes them particularly interesting from 
the point of view of the stellar evolution, stellar explosions, and the 
stellar 
nucleosynthesis. The kinetic energy of stellar ejecta is dissipated in
collisionless shocks, heating ejecta and the ambient medium to X-ray 
emitting temperatures. Through studies of X-ray spectra and X-ray 
morphology of  young SNRs, we can learn much about their progenitors and 
the physics of strong collisionless shocks.

\subsection{Recent Observations}\label{sec:yr_obs}

\subsubsection{X-Rays}

We are witnessing a rapid advance in X-ray observations of SNRs. Data
from the {\it ASCA}
and {\it ROSAT} satellites, much in evidence during the workshop,
have provided us with superb X-ray imaging and
with spatially-resolved X-ray spectra. The rapid progress in this field 
was
reviewed by {\it Hughes}. One of the most exciting prospects is the 
possibility of learning about SN progenitors from X-ray spectra of SNRs. 
For example, Hughes \&\ Singh (1994) found a massive ($\sim 25 
M_{\odot}$) SN progenitor in G292.0+1.8 by analyzing its {\it Einstein} 
SSS spectrum, which is dominated  by products of nuclear burning in 
massive stars such as 
O, Ne, Mg, Si, and S. Remnants of Type Ia SNe exhibit an entirely
different spectrum, with particularly strong lines of Fe, abundant Si, 
S, and 
Ar, but weak O and Ne lines. Such spectra were revealed by {\it ASCA} in 
the Balmer-dominated remnants in the  LMC (Hughes \etal~1995). However, 
detailed studies of the chemical composition are in general difficult 
for a variety of reasons. One of them is the nonhomogeneous (stratified) 
nature of stellar ejecta, as demonstrated by different physical 
conditions deduced for different elements in many young SNRs.

X-ray observations of a number of young SNRs were presented and 
discussed 
during the meeting. {\it Vink, Kaastra, \&\ Bleeker} showed puzzling 
results 
of their analysis of X-ray spectra of RCW~86, indicating low abundances 
of heavy elements and low ionization timescales. 3C397 ({\it Dyer \&\ 
Reynolds}) and RCW~103 ({\it Gotthelf, Hwang, \&\ Petre};
(Gotthelf, Petre, \&\ Hwang 1997) seem to have 
compact X-ray sources at their centers, possibly neutron stars which 
have not been detected at radio wavelengths. {\it Hwang \&\  Gotthelf}
considered a possibility that the Fe K$\alpha$ emission line in Tycho's 
SNR
is produced by the fluorescent emission from dust grains, but the large 
mass 
of dust required in this model might be in conflict with infrared data. 
{\it Hughes} presented X-ray measurements of proper motions in this 
remnant
(with {\it ROSAT}), finding fast knots with peculiar composition, 
presumably
clumps of stellar ejecta moving faster than the bulk of the X-ray 
emitting 
material. The {\it ASCA} spectrum of Kepler's SNR, another historical 
remnant, shows that its SN progenitor was most likely a massive star 
({\it Decourchelle, Kinugasa, \&\ Tsunemi}).

Perhaps the most exciting observational result is the accumulating 
evidence 
for the presence of nonthermal X-ray emission in young SNRs. 
Observations 
of Cas~A with {\it XTE} ({\it Allen}; (Allen et al. 1997)), {\it SAX} ({\it Vink}; Favata et al. 1997), and {\it 
GRO} (The \etal~1996) showed the presence of a hard high-energy tail 
extending to 100~keV. This is the second young SNR (after SN~1006, 
Koyama \etal~1995) where such nonthermal emission was unambiguously 
detected. As reviewed by {\it Keohane}, nonthermal  X-ray emission is 
most likely present in a number 
of young remnants, and has been also detected in older remnants such
as IC~443. This nonthermal emission might be produced either by 
nonthermal bremsstrahlung or by synchrotron emission of electrons with 
TeV energies as suggested by {\it Reynolds}, with the latter occurring 
in SN~1006 (Reynolds 1996). See Section~\ref{sec:accel} for more 
discussion.

\subsubsection{Infrared}

A new spectral window on SNRs was opened by the {\it Infrared Space 
Observatory}. Observations of the northern shell of Cas~A in the mid-
infrared (IR) with {\it ISOPHOT-S} revealed strong 
lines of singly and doubly ionized Ar and triply ionized S ({\it 
Tuffs}). 
Expansion velocities of several thousands of km~s$^{-1}$ were measured 
for this material, which must consist of undecelerated SN ejecta seen 
also in the optical. Infrared continuum was detected as well. The short 
wavelength (6 $\mu$m) continuum correlates spatially with radio 
emission, suggesting that the synchrotron emission from relativistic 
electrons might have been detected in the IR. Continuum emission from 
dust was detected at longer wavelengths with {\it ISOPHOT-S (Tuffs)} and 
{\it ISOCAM} (Lagage
\etal~1996). This emission correlates spatially with the undecelerated 
ejecta.
Observations at long (100 $\mu$m) wavelengths revealed
cold dust in the remnant's interior, presumably located in the unshocked
stellar ejecta. These important results herald a new era in IR
observations of SNRs.

\subsubsection{Optical and Radio}

Optical and radio observations, the traditional means of observing SNRs,
continue to yield important results. Optical
observations of nonradiative shocks in young SNRs, reviewed by {\it 
Raymond}, provide us with the crucial information about the velocity of 
the blast wave and its precise location. These nonradiative, Balmer-
dominated shocks are sometimes seen along the whole circumference of an 
SNR, e.~g., in SN~1006 and in RCW~86 ({\it Smith}). Optical and 
ultraviolet spectroscopy of these shocks gives us a unique opportunity 
to study physical processes taking place in collisionless shocks. In 
particular, UV observations
of SN~1006 indicate that only a small fraction of energy dissipated at 
the 
collisionless shock front is transferred to electrons at the shock front
(Laming \etal~1996). Optical observations of radiative shocks 
are also crucial for understanding the complicated dynamics and 
structure of 
young SNRs. {\it Sollerman \&\ Lundqvist} presented new high spatial 
resolution
observations of the oxygen-rich SNR 0540-69.3 in the LMC, finding 
strikingly different morphologies in emission lines of different 
elements.

There should still be a number of obscured, undiscovered
young SNRs in our Galaxy, as shown by analysis of radio 
survey data ({\it D. Green}).
Because our Galaxy is mostly transparent to both radio and X-rays, such 
remnants can be most effectively studied by combination of X-ray and 
radio 
observations, as demonstrated by {\it Dyer \&\ Reynolds} for 3C397. High 
spatial resolution of radio observations makes it possible to study 
young SNRs in nearby galaxies. {\it Muxlow, Wills, \&\ Pedlar} presented 
an analysis of over 30 young SNRs in the starburst galaxy M82, based on 
VLA and MERLIN observations. Many of these SNRs are apparently located 
within or behind dense HII regions, as shown by strong
free-free absorption at low frequencies.

\subsection{Hydrodynamics}

The morphology and dynamics of young SNRs depend on the 
distribution of the ambient medium and on the structure of stellar 
ejecta,
although in all models there is an outer shock (blast wave) propagating 
into
the ambient medium and an inner (reverse) shock propagating into the 
stellar
ejecta. In the absence of characteristic scales in stellar ejecta and
in the ambient medium, self-similar, spherically-symmetric solutions 
exist 
(Chevalier 1982a) and they
are widely used to interpret observational data on young SNRs. However, 
hydrodynamical instabilities and characteristic scales are
frequently present, necessitating the use of hydrodynamical (HD) or 
magneto-hydrodynamical (MHD) models. The current state of 
hydrodynamical modeling was reviewed by {\it Jones} in the context of 
young SNRs, and by {\it Norman} from a computational point of view. The 
most important conclusion is that it is now
feasible to model SNRs in one, two, and three dimensions even using 
some ``public-domain'' HD and MHD codes.
As a result, hydrodynamical modeling is becoming more common, which was 
reflected in the scientific results presented at the meeting. 
{\it Norman} talked about the importance and pitfalls of computational
methods.  The exponential growth in computational power and greatly  
enhanced
numerical algorithms has been the key to the improved role of 
numerical simulations for SNR studies.  Noting that computational power 
has consistently doubled approximately every 1.5 years, a rate which 
will be maintained into the next decade, Norman argued that we would 
soon finally have
enough power to attack complex simulations with fully 3-D codes
encompassing ``all the physics''.  He also warned not to over-interpret
the details of simplified simulations, and showed that often the
results from simple 1-D or 2-D cartoon models were later shown to
differ significantly from more sophisticated calculations.  The key is
the often unanticipated behavior of instabilities (\eg
Rayleigh-Taylor, Kelvin-Helmholtz), and he showed examples where they
severely compromised predictions of early models.  The warning that
subtle details of the physics can significantly change the turbulent
properties of the fluid was sobering in the context of the many papers
presented here that depended on the {\it details} of simplified
numerical simulations.

One-dimensional simulations are now routine. {\it Dwarkadas
\&\ Chevalier} considered an exponential density profile for ejecta of 
Type Ia
SNe, consistent with current models of the white dwarf explosions,
and found that the resulting structure and dynamics of their remnants 
are
very different from the standard self-similar solutions. This result 
should
serve as a reminder that hydrodynamical models of young SNRs are 
sensitive to 
the structure of ejecta, which might be rather complicated in detail as 
demonstrated by inspection of numerical models of SN explosions.
The density distribution of the ambient circumstellar and interstellar 
medium 
is also very important. This distribution is expected to be nonuniform 
around
massive SN progenitors because of the significant mass loss and the 
dynamical effects of the stellar winds. Dense circumstellar material, 
such as seen in Cas~A and Kepler's SNR, will strongly affect SNR 
hydrodynamics (Chevalier \&\ Liang 1989). 

Two-dimensional HD and MHD modeling of SNRs
is becoming standard. Such modeling should be generally preferred
over one-dimensional methods for 
young SNRs, because the contact discontinuity separating the shocked 
ejecta and
the shocked ambient medium is unstable (\eg, Chevalier, Blondin, \&\ 
Emmering 1992). The dynamics of the contact discontinuity was studied by 
{\it Dohm-Palmer \&\ Jones} from the initial free-expansion phase into 
the Sedov phase, when the swept mass is much larger than the ejecta 
mass. They
showed how to relate the evolutionary stage of the remnant, 
characterized by the ratio of the swept mass to the ejecta mass, to 
expansion 
rates measured at different locations within the SNR. Large-scale 
density 
gradients in the ambient medium will result in asymmetric, nonradial 
flows
which might be more readily observed than distortions of SNRs from 
spherical shapes (Dohm-Palmer \&\ Jones 1996).  MHD simulations of a SN 
explosion in a uniform ambient medium threaded by a uniform magnetic 
field show strong field amplification at the contact discontinuity at 
the magnetic equator, which may lead to formation of a barrel-shaped 
remnant, a point made by {\it Jung, Jun, Choe \& Jones}. 
Fragmentation of ejecta during the  
explosion and clumpy stellar outflows or an inhomogeneous ISM also  
necessitates use of hydrodynamical modeling. Two-dimensional simulations 
of an interaction of a young SNR with an interstellar cloud of
comparable size were presented by {\it Jun \&\ Jones},
who developed a model for the radio emission in multi-dimensions
based on simplified diffusive shock acceleration of the electrons.  {\it
Miniati \& Jones } presented simulation results for cloud-cloud 
collisions,
expected in the aftermath of supernova explosions, and found that these
often result in rapid destruction of the clouds. At a much later, 
radiative, stage in the SNR evolution, the formation of a thin, cool 
shell also leads to violent instabilities. {\it Wright, Blondin, 
Borkowski \&\ Reynolds} presented one- and two-dimensional simulations 
of these instabilities, finding particularly  vigorous instabilities and 
strong distortions of the shock front for SNRs expanding into a dense 
ISM.

Three-dimensional MHD simulations have been recently employed to model 
radio emission from young SNRs (Jun \&\ Norman 1996). Such large-scale 
supercomputer calculations open new frontiers in the SNR research. It is 
now
obvious that hydrodynamical simulations have become a powerful tool for 
theoretical studies of SNRs.

\subsection{Current Problems}

The most difficult task at present is to relate hydrodynamical modeling 
to
observations. A few of the observables, such as expansion rates and 
thicknesses of the flow structures, can be relatively easily determined from the models. 
However,
modeling radio and X-ray emission is in general difficult. {\it Jun \&\ 
Jones}
 and {\it Jung, Jun, Choe \& Jones} showed how multi-dimensional simulations 
can be used to model radio emission. These efforts are just the 
beginning, because we are still lacking the understanding of how 
electrons are accelerated in the SNR shocks. Very similar difficulties 
are encountered in modeling nonthermal X-rays. Thermal X-ray spectra are 
in principle easier to model, but in practice the  difficulties are 
formidable. The reason for these difficulties is our poor understanding 
of a number of topics, such as the 
amount of electron heating in collisionless shocks, the detailed 
structure of 
SN ejecta, their clumping, the presence of the nonhomogeneous 
circumstellar  medium, and the presence of dust. 

The difficult task of interpreting observations with the help of 
hydrodynamical models is perhaps best illustrated by Cas~A, the youngest 
SNR in our Galaxy. This is a remnant of a massive star explosion and a 
classic prototype shell SNR. It has been detected throughout the whole 
electromagnetic spectrum. Numerous optical and radio studies, reviewed 
by {\it Fesen} and by {\it Koralesky}, have provided us with a wealth of 
information about this SNR. However, as strongly emphasized by {\it 
Rudnick} in an overview to the Cas~A session, we still do not understand 
this remnant. The morphology of Cas~A at X-ray and radio wavelengths is 
dominated by a bright ring, rimmed by a fainter  plateau on its 
exterior, and a NW ``jet''. In addition to these large-scale features, 
there is also a lot of fine structure, well studied in the radio. What 
are relationships between all these features and the remnant's
hydrodynamics? Observations in various wavelength bands probe very 
different components of the remnant: synchrotron radio emission gives us 
information about relativistic electrons, thermal X-ray emission is 
produced by the bulk of the shocked hot gas, much cooler gas in 
radiative shocks emits at optical  wavelengths, and observations in 
infrared reveal still cooler gas and dust. What are the interactions and 
relationships between these various 
temperature components? There are also multiple kinematic components 
within the remnant. Fast-Moving Knots and Fast-Moving Flocculi seen in 
the optical are thought to be undecelerated, dense fragments of SN 
ejecta, expanding from the explosion center with velocities of several 
thousand km s$^{-1}$. Much slower (several hundred km s$^{-1}$) Quasi 
Stationary Flocculi are clearly shocked and accelerated remnants of 
stellar material ejected by the SN progenitor prior to the explosion. 
The
bright radio shell expands with intermediate velocities.
What is the self-consistent hydrodynamical picture for these kinematic
components? We do not know answers to most of these questions.

There is no consensus about the nature of the Cas A bright ring itself. 
Is 
this ring made from the shocked stellar ejecta, is it a shocked 
circumstellar shell, or is it both? Chevalier \&\ Liang (1989) noted 
that the slow expansion age measured for Cas A radio-emitting material 
($\sim 900$ yr; Anderson \&
Rudnick 1995) is consistent with the presence of the dense CSM. But Cas A 
might be also interacting with a molecular cloud in the west (Keohane, 
Rudnick,
\&\ Anderson 1996). Detailed studies of the ring kinematics along its
circumference and its relationship with the intensity, polarization, and 
the 
spectral index of the radio emission,
presented by {\it Koralesky \& Rudnick}, revealed large variations in 
the 
expansion of the ring, with more slowly expanding regions decreasing 
most rapidly in the  radio intensity. It is through such in-depth 
studies that we might finally learn about the nature of the ring and 
about the origin of its large- and small-scale 
asymmetries.

Modeling of its thermal X-ray emission is ambiguous in the absence of a 
commonly accepted dynamical model for Cas~A. It is then not surprising 
that the two existing models of Cas A's X-ray thermal spectrum are very 
different 
(Vink \etal~1996, Borkowski \etal~1996). However, both models require
that most of the shocked material consists of the swept-up ambient 
medium, 
although the X-ray spectrum is dominated by strong line emission 
produced 
by shocked stellar ejecta. Fe is notably absent in the shocked ejecta. 
{\it Hwang} presented an observer's view on the X-ray properties of 
Cas~A, 
through comparison with other young SNRs. The most outstanding features 
of
Cas~A are large ($\pm 1000$ km s$^{-1}$) emission-line Doppler shifts 
(Holt \etal 1994) and a good correlation between X-rays and radio 
(Keohane \etal~1996). Cas~A is also very bright in the optical. These 
might be characteristics of young remnants of massive stars, at least 
those 
which experienced extensive mass loss prior to the explosion.

The presence of strong lines of elements such as Si and S in the X-ray 
spectrum of Cas A means that we are observing heavy-element enriched 
material produced deep within the exploded star. As mentioned above, 
this offers us a unique opportunity literally to look into the stellar 
interior. This tantalizing possibility has always been the magnet for 
studies of young SNR, but the poor quality of observational data 
hampered progress in this field. This situation has changed with the 
launch of {\it ASCA}, and further 
improvement is expected after {\it AXAF} becomes operational. In order 
to take advantage of these new observations, we need to intensify our 
efforts to understand the dynamics and structure of Cas A and 
other young SNRs.

\section{Mature SNRs}\label{sec:mature}

Mature SNRs comprise the widest variety of remnants, exhibiting both
Sedov and/or radiative phase shell-type emission, often in the same
remnant.  {\it Dickel}, in an overview, emphasized this by showing
a gallery of multi-wavelength SNR images revealing objects with a
dazzling array of observable properties. These included remnants
driving both radiative and non-radiative shocks, and those displaying
classic shell as well as ``blow-out'' morphologies, the latter
reflecting interactions with an inhomogeneous ISM. The remnants also
displayed a wide range of magnetic field structure and polarization
intensities. He emphasized how SNRs control and are controlled by their
environments through kinematic interactions, heating, and excitation, a
theme reflected in many of the following presentations. Dickel also
emphasized that a comprehensive understanding of SNR evolution is only
possible through multi-wavelength studies.

\subsection{Individual Mature Remnants}\label{sec:talks}

{\it Levenson \& Graham} presented data on the Cygnus Loop 
with an emphasis on interpretation of this remnant as a cavity 
explosion (Levenson et al.~1997).  
Levenson showed an X-ray mosaic of many {\it ROSAT HRI} pointings and 
compared them to narrow-band optical (H$\alpha$, OIII, SII) images, 
radio continuum
images, and tracers of the neutral gas (\eg CO, HI).  The nearly
circular shell of the SNR as delineated in low resolution radio and
X-ray observations indicate that it has been expanding into a
relatively homogeneous region of the ISM.  However, in addition to a
smooth shell, the high resolution optical and X-ray data clearly
indicate the presence of many radiative shocks where the blast wave has
recently encountered large ($\sim$10 pc) clumps of dense ISM material.
Radial profiles of bright X-ray emission in these regions show
double-peaked structures interpreted as reflected shocks.  In contrast,
much fainter filaments exhibiting Balmer-line emission seem to trace
much faster, collisionless shocks where the blast wave is still
expanding into relatively low density material. This placed an
observational context underneath {\it Raymond's} previously mentioned 
theoretical discussion of non-radiative Balmer-dominated shocks 
penetrating partially neutral material.  Shull noted that Levenson's 
interpretation agreed with his previous kinematic and proper-motion 
H$\alpha$ survey of the Cygnus
Loop (Shull and Hippelein 1991).

{\it Leahy \& Roger} presented 408 and 1420 MHz
observations of the Cygnus Loop, highlighting the excellent surface
brightness sensitivity of the DRAO synthesis array and new polarimetry
instrumentation.  The degree of polarization across the source varies
from 39$\%$ to 2.4$\%$, with an inverse correlation between the degree
of polarization and X-ray intensity determined from {\it ROSAT} and {\it 
ASCA}.  The X-rays trace thermal material that can effectively
depolarize the radio emission. This provides a natural explanation
for the relationship, which provides a useful tracer of a mixed thermal 
electron population.

{\it Decourchelle, Sauvageot \& Tsunemi} combined
analyses of {\it ASCA} and {\it ROSAT} observations of the Cygnus Loop,
incorporating non-equilibrium ionization models.  They determined the
temperature, abundances and ionization structure from the northern rim
to the center. The inferred double-shell structure complements the
predictions of the SNR evolution calculations presented by
{\it Decourchelle}. The derived temperatures were higher than from
equilibrium ionization models. Trends in the radial
temperature and abundance profiles were also discussed.  {\it
Decourchelle, Sauvageot \& Bohigas} presented [Ne V] Fabry-Perot and
imaging observations of the Cygnus Loop.  This is the first time the
[Ne V] emission, which samples a medium close to the [O IV] gas, has
been mapped on a large scale in an SNR with the benefits of optical
spectral resolution and the convenience of ground-based observations.
The kinematics of the gas differs from the eastern to the western rim
of the remnant. This was interpreted in terms of the density structure
(inhomogeneous vs. homogeneous, respectively) of the emitting region.
Non-equilibrium ionization and temperature conditions required to
produce the [NE V] line were also discussed, as well as the effects of
evaporation and dynamic mixing with clouds of different filling factors
and sizes.  They also presented direct imaging data in the [Ne V] and
[O III] and correlated these data with X-ray emission from {\it ROSAT}.

A lively discussion around IC443 was introduced by {\it Strom}.  He
noted that Fesen's (1984) optical observations of filaments offset from
the ``classic'' IC443 shell-type SNR had motivated Braun and Strom's
(1986) Westerbork radio continuum study of the region.  Fesen's spectra
had suggested shock excited emission, and Braun and Strom's sensitive
radio images showed evidence for a separate and larger {\it nonthermal}
radio shell superimposed on the classic IC443
remnant, but offset to the east . 
Their images also suggested that the ``classic'' IC443
remnant may itself consist of two separate shells, so the complex
could consist of three or more separate remnants. Asaoka and
Aschenbach (1994) found with {\it ROSAT} a spherical, low surface
brightness, thermal X-ray source coincident with the Westerbork 92 cm
``second'' SNR.  Differing X-ray absorption characteristics for the
``second SNR'' and IC443 proper capped off their identification of the
new remnant G189.6+3.3. Much of the IC443 discussion at this workshop
centered on the continuing controversy over G189.6+3.3 interpreted as a
second SNR or as part of a single remnant expanding into a complex,
asymmetric cavity.

{\it Petre and Rho} reviewed the current IC443 X-ray data.  They noted
that the Asaoka and Aschenbach {\it ROSAT} PSPC all-sky survey exposure 
was
only a few hundred seconds, while data that Petre and Rho have since
analyzed include a 20,000 second exposure from Hester's pointed {\it 
ROSAT}
observations.  
Petre showed an HRI image clearly revealing the ``second'' SNR.  It now
appears to have a complex morphology, however.  Asaoka and Aschenbach
had based their identification of the new ``foreground'' SNR on the
need for two-temperature spectral fits to the X-ray data.  The deeper, 
pointed 
{\it ROSAT} data require only a single temperature,
but still require two different absorbing column densities. Thus, Petre
noted that while the simple picture of two separate SNRs with different
temperatures had been eroded somewhat, the basic Asaoka and Aschenbach
interpretation that the lower surface brightness emission originates in
front of IC443 is still valid. Despite this conclusion he was reluctant
to view the X-ray data as compelling evidence for or against either the
``two separate SNRs'' or ``one SNR in a complex region'' scenario.
However, he noted in favor of the former that at a presumed distance of
$\sim$1.5 kpc and age of $\sim$1000 yrs, it would be difficult under
the one SNR scenario to explode a star in the location of IC443 proper
and have X-ray emission extend as far to the east as observed.
Sustained shock velocities in excess of $\sim$ 30,000 km sec$^{-1}$
would be required.

{\it Hester} offered qualitative arguments {\it against} the
multi-remnant scenario for IC443. He noted that detailed observations
now reveal increasing numbers of remnants with complex morphologies,
including many multi-shell structures in presumed single-remnant
systems.  Referencing Levenson's Cygnus Loop talk he noted that in most
cases the complex morphology could be traced to the pre-existing
detailed structure in the ISM, usually some sort of cavity then
``lit-up'' by a single explosion event. Thus, he felt it was arbitrary
to single out IC443 as multiple remnants.  Hester argued that ``both''
remnants in the IC443 complex were interacting with the same HII
region/molecular cloud complex, indicating that the two objects were at
least physically close in space. He suggested it was possible to be
seeing the eastern bubble (a.k.a. the new SNR) ``in front'' of the
western bubble (aka IC443 proper), without proving that the two bubbles
were unconnected. He showed how this interpretation would change if
viewed from a different perspective, and argued that it seemed unlikely
to find two relatively high latitude, anti-center remnants in nearly
the same direction in the sky.  {\it Rho} disagreed and offered a quick
calculation indicating this could reasonably occur within a single OB
association.  Hester also illustrated how
arguments for multiple IC443 remnants could also be made for multiple
Cygnus Loop remnants, or in the case of Vela, for ``dozens of
remnants''.

{\it Leahy, Roger, \& Weimer} presented DRAO radio
continuum observations of IC443. Leahy's high dynamic range, high
surface brightness sensitivity 408 and 1420 MHz images were consistent
with Braun and Strom's early Westerbork work. The new maps show
evidence for the ``second'', low surface brightness eastern SNR only at
the lower frequency, confirming the nonthermal Westerbork
interpretation. However, Leahy maintained that an arc of emission to
the north-east of the remnant (which might correspond to part of the
2nd SNR in the Asaoka and Aschenbach interpretation) was {\it thermal},
possibly relating to the HII region.  His polarization data also
indicated strong Faraday depolarization within IC443 proper, providing
evidence for a mixed thermal electron population.  This high surface
brightness data, after a thorough analysis is completed (including
currently unprocessed HI data), might offer useful clues towards 
resolving the
IC443 controversy.  The relevance of the number of remnants associated
with IC443 to any fundamental question related SNR physics is this:
What validity can we attach to the statistical interpretations of all
of our surveys and catalogs of more distant objects, if we cannot even
determine the fundamental nature of the closest and largest objects?
These discussions also emphasized that while the lessons learned from
studying the nearest and largest ``pathological'' remnants like the
Cygnus Loop and IC443 are not easily transferred to other objects in
different environments and subject to different initial conditions, the
basic physical lessons we learn presumably are.  Hence detailed studies
of the ``micro-physics'' (eg, shocks) in nearby sources will always be
important and simply cannot be matched by observations of more distant
objects.

{\it Pineault, Landecker, Swerdlyk, \& Reich} presented 
DRAO (1420 and 408 MHz) and Effelsberg (1420 MHz) observations of
the SNR CTA1 (Pineault et al. 1997). These reveal evidence for spectral index
variations across the remnant, an issue discussed in several contexts
during the meeting and outlined in \S \ref{sec:accel}.
Pineault also noted a low-level extension in the radio
emission beyond the southern sharp rim of the remnant. This was
explained in terms of diffusion of electrons upstream of the shock with
a mean free path $\leq$0.02 pc, a result in agreement with theoretical
predictions (Reynolds 1994, Achterberg \etal 1994).
  
{\it Lozinskaya, Goss, Silchenko \& Helfand} presented 
radio, X-ray, and optical observations of S8, the only known SNR in the
galaxy IC1613. This SNR is a very bright X-ray and optical source, an
indication that it is embedded and interacting with a high density
environment, similar to the LMC remnant N49. The SNR's
appearance at the periphery of a complex of expanding and overlapping
shell-like structures supports this view.  
VLA multi-frequency data confirmed the nonthermal nature of the
radio spectrum first deduced by Dickel \etal  (1985), and established
the shell-like radio spectral index $\alpha \sim$-0.6.
Deep narrow-band optical images in a variety of lines ([OI], S[II],
[FeII], [NII], O[III], H$\alpha$) are used to constrain the kinematic
properties of the line emitting gas, and to determine \eg the radial
velocity of the SNR with respect to the mean velocity of the galaxy.

S8 was originally believed
to be an older (age $\sim$2x10$^4$ yrs) SNR in the late adiabatic or
radiative phase (D'Odorico and Dopita 1983, Peimbert \etal 1988), but
the new X-ray observations suggested instead that it may be a much
younger (Sedov,  age $\sim$3x10$^3$ yrs) cavity explosion SNR just
encountering the walls of a pre-existing cavity or a dense
cloud.  None of the existing models of shock excitation fit the measured 
line
ratios towards the SNR well, indicating that no simple single-shock
model is adequate, and that at least two shocks, one fast and one slow,
are required. This is not surprising in light of the complex ISM into
which the remnant is apparently expanding. In this context the {\it 
ROSAT} X-ray
observations, which barely resolved the source at $\sim$3", are 
difficult to
interpret, and higher angular resolution data are required for a
comprehensive analysis.  We were faced again with the common theme of
SNR evolution governed by the pre-existing structure of its ``birth-
place''
ISM.

\subsection{Surveys \& Systematics}\label{sec:mature_surv}

{\it Smith} discussed the importance of statistical studies (\eg
birthrates, distribution, energetics, etc.) of complete samples of SNRs
for exploring problems in stellar evolution, ISM structure, and for
increasing sample sizes of poorly understood SNR sub-classes.  This
began with the well-recognized limitations imposed on statistical
studies of Galactic SNRs by observational selection effects (\eg Green
1991). Smith also reviewed the common problems of obscuration and
distance uncertainties, emphasizing that the Galactic sample is
particularly incomplete in the youngest and oldest remnants.  The
importance of extragalactic SNR surveys received emphasis,
especially the obvious advantage of a common source distance. While
acknowledging the obvious tradeoffs in sensitivity and angular
resolution, particularly for radio and X-ray observations, the 
discussion
highlighted recent successes in nearby galaxies (in \eg, M82, M31, M33),
and the common feeling that advances in instrumentation were leading
extragalactic SNR surveys into a renaissance of sorts.  {\it Muxlow's} 
presentation on M82 illustrated this well.  Smith argued for the 
Magellanic Clouds as the best of both worlds, providing a population of 
remnants both at
the same distance and yet close enough to be examined in detail.
 
{\it Smith, \it Winkler, \& Chu} showed beautiful
optical (H$\alpha$, [SII], [OIII]) images from an ongoing Magellanic
Cloud Emission Line Survey (MCELS). A goal of the survey is to provide
a uniform set of flux calibrated emission-line images for all LMC SNRs,
and to find new remnants (seven identified so far) missed in previous
surveys.  They also illustrated the ability of the
[SII]/H$\alpha$ ratio to distinguish between SNRs and HII regions in
complex structures.  Another goal is to ``complete'' the LMC SNR sample
and make it available for statistical studies of evolution and
energetics, the merits of which were also discussed. The data are also
being correlated with X-ray and radio observations which greatly aid in
the identification of new SNRs (\eg Smith \etal 1994) and the analyses
of their properties.

{\it Williams} outlined a systematic study of LMC SNRs, a thesis
project with {\it Chu}.  The focus is to understand the evolution and
interaction of LMC SNRs with their surroundings through interpretation
of the latest X-ray, optical and radio observations. With a unique
co-distant, but relatively nearby sample of objects, the results of
this thesis are anxiously awaited, and Williams presented information 
(with
{\it Chu, Dickel, Smith, \& Milne}) on specific survey objects. They
examined the morphology and energetics of the ``breakout'' SNRs N11L
and N86 and their influence on the distribution of hot gas and
injection of heavy elements to the surrounding ISM, and noted that such
breakouts often allow us to probe otherwise unobservable low density
regions. The morphologies, velocity structures, and physical properties
of the remnant shells, postshock gas, and local ISM conditions were
characterized for both objects.  ({\it Leahy} and {\it
Pinealt} also addressed local ISM properties deduced from SNR
breakouts.) Williams and collaborators presented N44 as a
 showcase of specific remnant
physical properties which their data are being used to constrain.
Results on N44 include that the thermal pressure in the shell
($\sim$1.2-2.2 x 10$^{-11}$ dyne cm$^{-2}$) and cavity ($\sim$0.83-5.6
x 10$^{-11}$ dyne cm$^{-2}$) exceed the magnetic pressure ($\sim$2.3 x
10$^{-12}$ dyne cm$^{-2}$), and that the inferred kinetic energy
($\sim$2.4-4.5 x 10$^{49}$ ergs) is low and implies the remnant may not 
be
adiabatic. In addition, the size ($\sim$30 pc), X-ray temperature
($\sim$0.24-0.46 keV), and derived shell mass ($\sim$240-450 M$_\odot$)
and expansion velocity ($\sim$100 km s$^{-1}$) were consistent with the
derived age of $\sim$5.8-14 x 10$^4$ yrs. Work continues to constrain
parameters further from HI data and to make a comparison with the
nearby supershells N11L and DEML316.  When the analyzed LMC SNR 
sample
is large enough, it will constrain the distribution of total remnant
energies, establish trends in ratios of thermal to kinetic and magnetic
to relativistic energies, and delineate the structure (\eg density,
clumpiness, filling factors) of the ISM in the LMC as a whole. The work
of Duric \etal (1995) and Muxlow (discussed below) offers additional
examples of conclusions drawn from meaningful samples of co-distant,
extragalactic SNRs.  These include the efficiency of particle
acceleration in M33 and the ISM density in M82, respectively.

{\it Wallace} presented examples of SNRs interacting with
HI and H$_2$ clouds, including IC443, CTB109, HC40, W44, and W51C.  He
noted SNR physical properties that could be constrained indirectly from
such observations:  1) SNR kinematic distances from velocities of
associated HI/CO features, or possibly from shock excited masers
discovered by Frail (\eg see Frail \etal 1996) and presented by 
{\it Goss};  2) SNR shock velocities from the
measured velocity of an expanding shell of recombined material; 3)
pre-shock densities of clouds SNRs are expanding into from measurements
of multiple molecular lines. Again note that masers can
now serve as diagnostics of the post-shock gas, including utilizing
Zeeman splitting to determine magnetic field strengths.  Wallace
used W44 to illustrate an exotic example of SNR/cloud interactions.  Koo 
and
Heiles (1995) had detected a disrupted HI shell interior to the
radio shell in W44, and interpreted it as a pre-SN wind-swept bubble
subsequently overtaken by the SN shock.

{\it Decourchelle \& \it Chi\`eze} presented theoretical results of
mass density radial profile calculations for SNRs evolving from
different mass progenitors with and without stellar winds.  They
assumed spherical symmetry, adiabatic expansion, a homogeneous ambient
medium (no clouds), no thermal conduction, and  power-law profile
ejecta with a flat density core. The calculations were aimed at
predicting observable features in real X-ray data so they adopted the
Sedov model explosion energy and ISM density parameters for the Cygnus
Loop (from Ku \etal 1984). Radial profiles of mass density for a
variety of ages were shown for the different initial starting
conditions.  Profiles of the projected value of the density squared
(proportional to the X-ray flux) and of temperature were also
given to connect to real Cygnus Loop data. The validity of their
calculations extended to longer time-scales than previous self-similar
calculations (\eg Chevalier 1982a).  The variety of multi-shell and
reverse shock signatures which appeared in the profiles were striking,
in contrast to the smooth, late-time Sedov (no ejecta, no stellar wind)
profiles also shown. For example, a massive progenitor with a red
supergiant wind generates three separate mass shells, an inner one for
the shocked ejecta, a middle one for the shocked stellar wind, and the
outermost for the shocked ISM.  The SNR blast from a massive progenitor
with a blue supergiant wind encounters a dense ISM shell. In that case
the shocked ISM shell dominates over the shocked ejecta and shocked
stellar wind signatures.  Decourchelle showed observed Cygnus Loop PSPC
and {\it ASCA} count profiles displaying structures remarkably similar 
to the
multi-shell features revealed in some of these model calculations.  She
also reported ongoing work on deriving spectra incorporating
non-ionization equilibrium calculations in the hope that these can also
serve as indicators of progenitor mass, ambient medium density (\eg
wind vs. no wind), and evolutionary stage.

The rich variety of prominent, non-Sedov features presented by {\it 
Decourchelle} suggest that modern X-ray measurements could serve as 
powerful SNR evolutionary diagnostics. However, {\it Hughes} noted that 
despite hard 
effort, he had failed to find observational evidence for ejecta in a 
mature
SNR.  Decourchelle emphasized that the best hope of finding such
signatures was in massive progenitors with strong stellar winds.
Perhaps the consequences of her assumptions together with point-spread
function and projection effect limitations make such structures harder
to observe than her calculations suggested. For example, {\it Shelton
\& \it Long} separately emphasized the important effects on X-ray
emission of thermal conduction or a clumpy ISM, respectively. On the
other hand, {\it Vink, Kaastra \& Bleeker} showed that
spectroscopic evidence exists for ejecta in the mature SNR RCW 86.
They presented both {\it ASCA} and {\it ROSAT} data on RCW86, identified 
as a
``cavity'' explosion from the low ambient ISM density ($\sim$0.1
cm$^{-3}$) determined from their analysis. They identified at least two
spatially and spectrally distinct regions with significantly different
temperatures (kT$\sim$0.8 keV and $>$ 3 keV) and showing strong
departures from ionization equilibrium (n$_e$t$\leq$300 cm$^{-3}$yr).
The Sedov analysis gave a distance of 2.8 kpc and age of 7000 years,
and they identified a possible association with a known OB association
at 2.5 kpc.

{\it Muxlow, \it Wills, \& Pedlar} presented recent results from a
continuing VLA/MERLIN study of radio SNRs in the starburst galaxy M82.
Muxlow \etal (1994) had identified over 40 discrete shell-type
sources as being SNRs, all resolved with the MERLIN 50 mas beam at 5 GHz 
(or 0.75 pc for an M82 distance of 3.2 Mpc). The
sources, all pre-Sedov, were found to be smaller and brighter (and thus
presumably younger) than the equivalent Galactic and LMC populations.
The sample was large enough to estimate the SN rate in M82 at
$\sim$0.05 per year and from the flux density versus diameter
statistics to infer ongoing relativistic particle acceleration.  Their
newest data extends the M82 study to a
larger and older population of SNRs.  Much older remnants were still
strongly selected against since sources larger than 5 pc blended into
the background. However, Muxlow did attribute the extended nuclear
radio emission in M82 to a large number of much older remnants and
determine that the component of this emission within $\sim$180 pc of
the core had a considerably steeper spectrum than the extended emission
at larger radii.  This steeper spectrum region lies interior to the
known molecular ring of material in M82 (Nakai \etal  1987).  An
unexpected result was that many of the remnants show low frequency
turnovers in their radio continuum spectra at relatively higher
frequencies than those seen towards galactic SNRs (Kassim 1989).  This
is attributed to free-free absorption by ionized gas in their immediate
surroundings with emission measures ($\sim$10$^6$~pc cm$^{-6}$),
comparable to properties of Galactic giant HII regions.  These
measurements place the discrete sources relative to the ionized
component of the M82 ISM. HI absorption and CO emission observations
are also being analyzed in order to fix the SNRs dynamically, relative
to the neutral gas and molecular clouds in M82.

A statistical analysis of the most compact sources yielded a Diameter
(D) vs. time (T) relation D$\propto$T$^{0.6}$, suggesting greater
deceleration than free-expansion (D$\propto$T) or models for very young
SNRs (D$\propto$T$^{0.8}$, Chevalier 1982b), but still less decelerated 
than
Sedov (D$\propto$T$^{0.4}$).
This implied that the expansion of even the largest and oldest remnants
was still dominated by ejecta, which sets an upper limit to the M82 ISM
density. Conversely the thermal absorption optical depths were used to
set a lower limit on this quantity, so that together they implied an M82
ISM density $\sim$30 cm$^{-3}$ and a filling factor of $\sim$0.1. This
compared favorably with radio recombination line results (Roelfsma \&
Goss 1992).

This impressive study, along with other recent surveys in
nearby galaxies (\eg M33 by Duric \etal 1995, the LMC studies
presented here), shows how much can be learned from co-distant samples
of SNRs in nearby, face-on galaxies. {\it Chu, Williams, Smith
and D. Green}  all made these same points. But
{\it Biermann} followed with the observation that the absorption, sensitivity,
and angular resolution constraints on these surveys ironically
re-introduce selection effects not unlike those which dog the Galactic
surveys. Hence as much care must be taken in interpreting the
statistical conclusions drawn from these extragalactic SNR surveys as
is appreciated for Galactic surveys.

{\it Kothes} emphasized means for inferring SN progenitor types from
present day mature SNR radio observations. He presented SNR evolution
model (\eg Gull 1973) plots of radio surface brightness vs. radius for
various SN progenitors and showed where some well known SNRs (\eg
Kepler, Tycho, SN1006, Cas A, W28) fell on these curves. He concluded
that when bright radio emission is seen at small radii it was difficult
deciding between SNIa or SNII progenitors, while when bright radio
emission is seen at large radii it implied a strong stellar wind
progenitor.  A caveat for the first case was that a maximum light
equivalent surface brightness resulted for assumed ambient densities of
$\sim$10 and $\sim$100 cm$^{-3}$ for the SNIa and SNII cases,
respectively, so that if an estimate of the density were also available
(\eg from X-rays) it could allow one to differentiate between these
types. This was used to infer an SNIa origin for Kepler, Tycho, and
SN1006.

For faint radio emission at large radii in evolved SNRs,  {\it Kothes} 
pointed 
to the X-ray emission as the clue to the progenitor identity.  If the 
SNR is
X-ray bright, implying the mean density in the shock is high, it was
probably a core-collapse progenitor with a strong stellar wind;
conversely if X-ray faint, it implied a SNIa progenitor in the inter-arm
medium. He emphasized the strong selection effect against identifying
mature Type Ia SNRs, since the white dwarf progenitors likely exploded
in low density inter-arm regions and have weak radio and X-ray
emission. While there has been some success in identifying young SNRs
from presumed SNIa progenitors from their Balmer dominated spectra,
Kothes was describing the physical conditions around which future
observations might best hope to identify mature SNIa remnants. 

{\it Kothes} also applied these lessons by presenting a nice example
of a newly identified and presumed mature SNIa remnant, G182.4+4.3.
First recognized from the anti-center region of the Effelsberg 11 cm
Galactic plane survey (F\"{u}rst \etal 1990) and subsequently
re-observed at 4.85 GHz, the identification as a shell-type SNR in the
adiabatic phase was based on the radio morphology, tangential magnetic
field structure, and radio spectrum ($\alpha \sim -0.4$). Kothes used
the radio surface brightness to infer a very low ambient medium density
($\sim$0.02 cm$^{-3}$) and together with the lack of any detectable
X-ray emission from {\it ROSAT}, to conclude that the progenitor was a 
SNIa
in the inter-arm region.
 
The SNR CTB87 was also discussed by Kothes, where the composite
remnant was shown to exhibit the classic steeper ($\alpha \sim-0.5$)
and flatter ($\alpha \sim-0.1$) radio spectra from its shell and
plerionic components, respectively.  Centrally peaked X-ray emission
was interpreted using the White and Long (1991) model, and together
with an estimate of the ambient medium density (from HI) he argued for
a core collapse explosion in a cloudy medium with a type SNIb/c
progenitor.  With additional assumptions, physical properties were then
derived for this SNR.

{\it Petre} displayed beautiful {\it ROSAT} HRI images of
IC443, Puppis A, W44, and Cas A among several other SNRs. The 5"
resolution images illustrated the impact the ${\it ROSAT}$ and ${\it 
ASCA}$ X-ray
``revolution'' has had on current SNR research. These images recalled
early VLA radio images, where for the first time objects known to
contain complex structure from simpler interferometer measurements were
finally revealed in their complete splendor. It has been argued that
not a great deal more has been learned about the fundamental nature of
radio galaxies since they were first resolved. The challenge to SNR
X-ray studies is to prove that the images shown in Petre's
poster will translate into a real increase in our physical knowledge
about SNRs. Many of the excellent papers presented at this workshop 
directly addressed this challenge.

{\it Gaensler \& A. Green} presented work on bilateral SNRs. They note 
that the morphology of this distinct
subset of composite and shell-type SNRs may be governed by either
``intrinsic'' or ``extrinsic'' effects. An example 
of a intrinsic morphological driver
is a toroidal distribution of ejecta in the SN explosion
(Kesteven and Caswell 1987), while one extrinsic explanation is
preferential electron acceleration when the shock normal and the
ambient magnetic field are perpendicular, also called
``quasi-perpendicular acceleration'' (Roger \etal  1988, Fulbright and
Reynolds 1990). From a sample of 17 well resolved ``barrels'' they found 
a
significant tendency for the bilateral axis to be aligned with
the Galactic plane, pointing to extrinsic effects as the morphological
drivers.  Since the Galactic magnetic field is oriented mainly along
the plane (Mathewson and Ford 1970, Ellis and Axon 1978) they argue for
compression of the ambient field lines and/or quasi-perpendicular
acceleration as driving the bilateral morphology of these SNRs. However
they argued that the ambient field also helps shape the wind-blown 
bubbles
of massive stars and the pre-SN ISM environment into tunnels,
interfaces and cavities elongated parallel to the plane (K\"{o}nigl
1982; Stone and Norman 1992), and that these effects may be required to
explain the asymmetric barrels with more complex structure (\eg
G320.4-01.2, G356.3-01.5, and G166.0+04.3). Then, viewing angle
selection effects and inhomogeneities in the local ISM and magnetic
field structures may account for why even $more$ SNRs do not display
bilateral morphology.

{\it Shull}  addressed the shortcomings of Woltjer's (1972) cartoon
of SNR evolution
and suggested we instead adopt a new ``flexible'' approach to SNR
classification. He identified the erroneous assumptions of uniformity
and symmetry as the main culprits for the failure of the old system.

Shull's cartoon illustration of a system to overcome these deficiencies
considered SNRs which might evolve from explosions of one of two
possible progenitor stars (O or B) in one of two possible environments
(symmetric or asymmetric). In this case four possible birth scenarios
were possible, and he described the observable differences in the four
cases. He offered the two LMC SNRs, N63A and N49, as cases of
``symmetric O-Star'' and ``symmetric B-star'' scenarios, respectively.
Here the observable properties may relate back to differences
in the pre-SN mass loss (winds) and ionization (\eg UV photon flux)
characteristics of the progenitor.  From birth-place scenarios 
specifying at 
least 
the progenitor type and ISM symmetry and
state, Shull suggested we could develop new terminology to specify the
SNR's evolutionary state within that scenario. The idea is interesting,
but even Shull admitted that there will always be remnants like the
Cygnus Loop, which is a half-cavity explosion, making it a ``mutt''.
It remains to be seen if we are brave enough to forge such new schemes
and lead the community away from the ``deficient'' yet nonetheless
still popular classifications that we have grown so comfortable with
over the years.

\section{Old Remnants and the Interstellar Medium}\label{old}

As SNRs disperse into the interstellar medium, they may be hard
to recognize as individual objects, especially in our own galaxy.
At the same time, galaxies contain many supershells 
caused by multiple supernovae
and stellar winds.  They are thought to dominate the morphology and
often form chimneys that connect to the halo, providing a direct
connection for gas and relativistic particles.  They profoundly affect
the ionization of the ISM, with their porosity allowing something like
half of the ionizing photons from a star cluster to escape the immediate
vicinity and travel long distances to provide a pervasive Warm Ionized
Medium.  These were some of the issues placed on the table by {\it 
McCray} and by {\it Heiles} in their reviews of old SNRs and the ISM.  
{\it Shelton, Cox \& Petre} also discussed how ancient SNRs in 
the galactic halo could be responsible for the hot (UV and X-ray
emitting) gas.  SNRs also push the gas and magnetic field around, 
forming 
huge
cavities with material and magnetic field lines pushed to the outside
walls.  The displaced matter exhibits itself in atomic, ionic, and
molecular lines, which can be observed even in external galaxies.

\subsection{Old Remnants and Bubbles}

{\it Chu} summarized some of the key findings from LMC studies. 
 The LMC, having a small, known distance, a small
foreground and internal extinction, and a nearly face-on view, allows us 
to identify the largest SNRs and to study the merging of SNRs with the
ISM.

     The LMC SNR 0450-709, with a size of 104x75 pc, is the largest
SNR known and is presumably very old (Mathewson \etal
1985).  This SNR should not be confused with superbubbles that 
exhibit SNR signatures, such as bright X-ray emission and 
nonthermal radio spectra.  The shell of 0450-709 is the SNR shell
itself, while the shell of a superbubble is shaped collectively
by stellar winds and supernova blasts from an entire OB association.
The SNR signatures of a superbubble are generated when a supernova
goes off near the superbubble shell walls (Chu \& Mac Low 1990).

     {\it Chu} noted that the LMC shows a large-scale distribution of 
hot ($10^6$ K) gas (Snowden \& Petre 1994).  Some is concentrated near 
star
forming complexes, while 
the rest exists in relatively quiescent regions.  The hot gas 
most likely represents a conglomerate of old SNRs.  Studies of this
hot gas as well as its underlying massive star content are underway. 

{\it R. Smith}, working with D. Cox, has
modeled the Local Bubble  using a one-dimensional
hydro code ({\sc odin}) that can simulate multiple supernova remnants,
with non-equilibrium ion evolution and dust.  Their model assumes that
the local interstellar medium was a cool ($10^4$\,K) gas approximately
5-10 Myr ago; it was then disturbed by 2 or 3 supernovae exploding
within 20-30 pc of each other over a period of 2-4 million years.  The
Local Bubble is the leftover hot gas from these explosions.  The model 
predicts the X-ray emission from such a bubble, as well as ionic 
abundances for hot gas ions such as O {\sc vi}.  These compare well with 
the soft X-ray data from the Wisconsin all-sky survey and the {\it 
ROSAT}
PSPC.

\subsection{Old Remnants and the Galactic Magnetic Field}

{\it Heiles} noted that although multiple supernovae also displace
the galactic magnetic field, this is not so easy to study, and it is
observable only in the nearby Galactic supershells.  Starlight
polarization is the best tracer of these fields, and
provides the plane-of-the-sky orientation (but not the
direction) weighted by extinction.

        Traditionally, the local supershells and supernova remnants are
defined by the four major Radio Loops.  However, except for Radio Loop 1
(the North Polar Spur or NPS), the others exhibit nothing other than
radio emission to make us think that they are indeed related to SNR.
Furthermore, another prominent supershell, the Eridanus Loop, exhibits
the expanding shell and diffuse X-ray emission we expect for a multiple
SNR event, but no significant diffuse radio emission.  Thus, the radio
loops are not reliable indicators of supernova shells.  Furthermore,
their traditional interpretation as limb-brightened shells doesn't agree
with various aspects of their morphology, \eg  with the ratio of shell
thickness to radius.

          {\it Heiles} concluded
 that the radio continuum loops are not very good
tracers of interstellar shell structures, nor are they limb-brightened
shells.  What are they? He believes that they are magnetic flux tubes
``lit up'' by an excess of relativistic electrons.  The NPS is
the best example.  The deformed field lines are obvious in the classical
Mathewson and Ford map of starlight polarization.  A careful look at
these shows a very good match---including not only general shape but
also sharp bends in the field lines---to a model of magnetic lines
deformed by an expanding shell centered near $(l, b) = (320^\circ,
5^\circ)$; this is very close to the center derived from the expanding
HI shell and is significantly different from the radio loop center $(l,
b) = (329^\circ \pm 1.5^\circ, 17.5^\circ \pm 3^\circ)$, illustrating
again that the radio continuum loop emission does not define the
physically important structure, which is the expanding supershell.

        In the NPS, the most intense radio emission arches up towards
positive latitudes near $l \sim 30^\circ$ and there are several roughly
concentric filaments of different radius.  The filaments lie roughly
parallel to the stellar polarization, again suggesting that the
filaments trace magnetic field lines.

          This pattern match, plus the many bright radio filaments that
exist not only near the periphery, suggests that the bright radio
filaments trace particular distorted magnetic field lines.  This is a
very strong indication that the brighter portions of Radio Loop 1 are
not bright because of limb brightening.  Rather, they are defined by
distorted field lines that happen to be ``lit up'' by relativistic
electrons.  Whatever the physical cause, the effect is huge: the mean
Galactic synchrotron emissivity near the Sun is $\sim 7$ K kpc$^{-1}$
and the bright filaments have at least several Kelvins pc$^{-1}$,
$\ge 500$ times higher! These bright field lines seem to run
preferentially close to dense interstellar clouds, such as the Ophiuchus
dark clouds; perhaps the interaction region that exists between
particular dense pockets of gas and the expanding shock are places where
relativistic electron generation occurs particularly efficiently.

\subsection{The Environment of SNRs and their Detectability}

{\it McCray} summarized many of the issues raised at
this conference which provided abundant confirmation of the important 
notion that the morphology and visibility of supernova remnants are 
determined largely by their circumstellar environments.  Since supernova 
remnants result from the impact of the supernova ejecta with 
circumstellar gas, the visibility of SNRs is highly biased in favor of 
those with massive progenitors, such as Cas A, which are concentrated in 
the disk of the Milky Way.  

Many young SNRs from massive progenitors are bright because the 
supernova ejecta are interacting with nearby gas expelled by the 
progenitor itself, presumably during a red supergiant stage.  This 
circumstellar gas is likely to have mass comparable to that of the 
supernova debris and will not extend much further than a few parsecs.  
After several centuries, the blast wave from the supernova will pass 
through this relatively dense circumstellar gas.  

But the interstellar medium beyond this relic red giant wind is also 
likely to 
have been highly disturbed by the progenitor evolution.  For example, 
the 
stellar wind from a blue supergiant stage preceding the red supergiant 
stage 
could have displaced the interstellar gas with an interstellar bubble of 
hot, 
low density gas surrounded by a dense shell of radius $\sim 20$ pc or 
more.  
If so, a supernova blast wave may remain nearly invisible for several 
thousand 
years, from the time it exits the relic red giant wind until it strikes 
the 
bubble wall.  When it does, we will see a ``mature'' supernova remnant, 
such 
as the Cygnus Loop ({\it N. Levinson}).  In such a scenario, the 
actual age of the SNR may be considerably less than the kinematical age 
estimated from the radius of the filaments divided by the expansion 
velocity.
Moreover, counts of supernova remnants as a function of age may have 
huge selection effects.  People should exercise extreme caution in 
inferring supernova rates from counts of mature and old SNRs.

Even more important, most massive stars are found in clusters.  
Therefore, 
most type II supernovae will not be the first one in the vicinity, but 
more likely will occur in a medium that has been highly disturbed by the 
action of 
previous supernovae.  The typical lifetime of a massive star that is 
likely to end as a supernova (a few $\times 10^7$ years) is not long 
enough for the interstellar medium to back-fill the cavity left by a 
previous supernova.  An OB association will give rise to several 
supernovae in this interval.  For example, even a relatively modest 
cluster such as the Pleiades should have already produced several 
supernovae.  In a cluster, each subsequent supernova will rejuvenate the 
cavity left by the previous ones, causing the formation of a 
``superbubble'' with diameter $\sim 50 - 100$ pc or more (McCray \& 
Kafatos 1987).  The superbubble interior may be quite irregular, 
containing high velocity filaments moving chaotically, as we see in the 
Vela-Puppis region.  Other prominent superbubbles in the Milky Way are 
those surrounding the Cygnus OB1 association, the Aquila supershell 
(Maciejewski \etal 1996), and the Monogem Ring (Plucinsky \etal 1996).  
We also see several superbubbles around OB associations in the Large 
Magellanic Cloud (Oey 1996).  

Most SNRs are likely to be found in superbubbles, but it may be 
difficult to 
identify the old ones individually because they have merged with other 
old 
SNRs.  We can only be sure to see the young ones, which are still 
interacting with circumstellar gas expelled by their progenitors.  Some 
LMC superbubbles are brighter than expected in soft X-rays (Chu \etal 
1993), and some are expanding faster than expected (Oey 1996).  Perhaps 
these phenomena can be explained as transients due to the impact of the 
most recent old SNR with the walls of the  superbubble (Chu \etal 1993).

\section{X-Ray Filled Composite Remnants}\label{sec:composite}

Much discussion at the workshop dealt with the origins and significance
of remnant morphology.
SNRs have been usually classified based on their 
\underbar {radio} morphology into three broad categories:
 shell--type,  centrally concentrated or ``plerions'', and composites.
{\bf Shell--type} remnants, which represent almost 80\% of the 215  SNRs
cataloged in our Galaxy, depict a hollow morphology in
radio wavelengths, with the flux density increasing from the center to
the periphery; the polarization is  generally weak (p $\simeq$ 5 to 15
\%), and the radio spectral index $\alpha$  
varies between  -0.3 and  -0.7.  The interaction of the
shock wave with the surrounding interstellar medium is responsible for
the radio emission (examples of this
kind are Tycho's SNR, SN 1006, Cassiopeia A, Cygnus Loop).
{\bf Plerionic} remnants include all the ``Crab--like'' sources, with a
filled-center appearance; \ie the flux density decreasing from the 
center
to the periphery. They are highly polarized (p $\simeq$ 20 to 30\%) 
and the spectrum is flat,
with a spectral index $\alpha \geq -0.3 $. In this case,
rotational energy losses from a central pulsar power the non-thermal
nebulae. Less than 10\% of the galactic SNRs belong to this class; 
examples of this class are, in addition to the Crab nebula,
3C58, MSH 11-54, etc.  {\bf Composites} are an intermediate class of 
SNRs 
which share characteristics of pure shell remnants and pure plerions; 
\ie a 
limb
brightened radio shell with steep spectrum and a central flat spectrum
radio nebula (\eg CTB 80, Vela XYZ, 
G 326.3-1.8).

In recent years, however, from the increasing number of SNRs surveyed
in the X-ray range with good angular and spectral resolution (mainly
with the {\it ASCA} and {\it ROSAT} satellites), the so-called 
``composite'' class
grew to include all types of {\bf centrally--influenced} remnants; \eg
objects with a shock wave powered \underbar {radio} shell, filled by 
centrally enhanced
\underbar {X--ray} emission. The X-ray radiation can consist of  a hard 
X--ray
compact nebula  (detected in the range $\sim$ 4 to 9 keV), non--thermal 
in 
nature, and/or extended soft  thermal X--ray emission (detected in the
spectral range $\sim$ 0.5 to 4 keV). 
Examples of this type are W44,  W28, 3C400.2, MSH 11-62,  VRO.42.05.01. 
A special discussion was organized to deal with the strong interest in 
resolving the nature of the subset of SNRs
now generically known as ``thermal X-ray  composites'', \ie , remnants with 
a
thermal X-ray bright center and a radio shell.

In dealing with these sources, we have to focus on two basic question: 
{\it 
i)}
Do these sources form a single homogeneous group, with common physical 
processes giving rise to
the observed morphologies in the different spectral ranges, and {\it 
ii)}
 What are the physical mechanisms responsible
for the observed characteristics? The subject is being actively studied 
both
from theoretical and multi-spectral observational approaches.

\subsection {Observational Results}\label{sec:composite_Obs}

\subsubsection {Imaging and spectral studies of centrally-influenced 
SNRs}

A comparative study of several members of the  ``thermal X-ray'' 
composites 
was
presented by {\it Rho}.  Central brightness X-ray enhancements of about 
2 - 5 times the brightness in the periphery are found in {\bf W44}, 
{\bf W28}, {\bf 3C391}, {\bf MSH11-61A}, 
and {\bf W63}, and as high as 5 - 13 times in {\bf IC443}.
 Temperature profiles were shown to be largely uniform across the
remnants, without radial dependence. Density and pressure profiles 
allowed
{\it Rho} to conclude that N$_{\rm H}$ variations are not significant 
enough 
to
change the center-filled morphology, and the appearance must have 
an intrinsic origin. 

Based on the global X-ray properties, {\it Rho} classifies the analyzed
SNRs as follows: W44, W28, 3C400.2, Kes 27, MSH11-61A, 3C391, and CTB1,
as belonging to the ``mixed-morphology'' type (M-type: thermal X-ray
emission inside hollow radio shells). Another four remnants: W51C, CTA1,
W63 and HB21 are classified as ``possible composites''. The SNRs W49 B
and 3C397 are probably not composites, and MSH 11-54 is found 
definitely not to belong to this class. The remnants IC443, Kes 79 and
HB3 are shown to be similar to M-composites, but show other dominant
physical phenomena which make the X-ray morphology complex.
{\it Rho} concludes that at least 8 to 11 \% of all cataloged galactic
SNRs belong to the M-type group. This corresponds to  over 25\% of all 
the X-ray
detected SNRs in our Galaxy. The primary mechanisms proposed to produce this 
mixed
morphology are evaporation of clouds, reflected  and reverse shocks. 
Also, the scale height from the galactic
plane was analyzed for the different classes of SNRs concluding that the
galactic height increases for the sequence Crab-like$\rightarrow$ M-type 
$\rightarrow$ shell-like remnants. 

{\it Burrows} reported {\it ASCA} {\it SIS} and {\it GIS} observations
of {\bf VRO 42.05.01} and {\bf 3C400.2} (Guo \&\ Burrows 1997). 
Both remnants have unusual shapes in radio
wavelengths suggestive of being the result of the breakout of a 
spherical
SNR into a lower density region. The X-ray radiation, thermal in nature,
fills the interior of both radio remnants. In  VRO 42.05.01 the X-rays
peak in the ``wing'' region,  while in   3C400.2 the X-ray emission 
attains a maximum to the NW, exactly at the intersection of the two 
circular shells that form the
radio remnant.  Although the two remnants have similar ages and
comparable morphologies, the X-ray spectra are found to be dramatically
different. In  VRO 42.05.01, the spectrum is consistent with thermal
emission but is nearly featureless, whereas in  3C400.2 the best 
spectral fit 
is obtained with a simple thermal bremsstrahlung with strong lines of Si 
and 
Mg. {\it Burrows }
suggests that the spectral differences are a consequence of abundance
differences in the interstellar environments of the remnants.

{\it Slane} investigated the SNRs {\bf MSH 11-61A} and {\bf W28}.
Two different models are discussed to interpret the {\it ASCA} X-ray 
data of 
these remnants: one scenario in which the shells have recently gone
radiative, thus leaving only the hot interior to persist in X-rays 
(using
a 1-D shock code employing simple radiative cooling and Coulomb
equilibration between electrons and ions), and
another which explains the central emission enhancement by the presence
of cool clouds that slowly evaporate in the hot remnant interior, using
 White and Long's (1991) similarity solution. For MSH 11-61A {\it Slane} 
finds
that this latter model can reproduce both the observed 
brightness distribution and the temperature profile. However the cloud 
parameters 
required by the solution appear physically problematic; 
evaporation timescales
of 50-100 times the age of the remnant are implied. Simulations 
with the shock code show, on the other hand, that the brightness profile 
for MSH 11-61A can be achieved for an old remnant in which the 
shell has recently gone radiative, assuming adequate physical parameters 
for
the SN and its environs.  In this case, however, it appears that 
the temperature profile drops more rapidly with radius than what is
 observed for this remnant. 
For W28, the observed brightness profile appears too 
steep (\ie , too centrally
peaked) for either of the models above to reproduce the observations. 
Attempts
with the cloudy ISM model require evaporation timescales which are 150 
times
the remnant's age (or larger), but even these over predict the brightness 
just
outside the bright core. 

 {\it Harrus} reported studies of {\bf MSH 11-62}. This SNR, unlike the
previous ones, has a  centrally-brightened morphology in the radio
band, with a flat
spectrum ($\alpha$ = -0.29), strongly polarized central  component 
surrounded
by a shell. {\it ASCA} observations have
allowed these investigators to unambiguously identify two distinct 
contributions to the
X-ray emission in the interior of the remnant: a thermal
extended (diameter $\sim 10^\prime$) component (at energies below 2 keV)
and a non-thermal point source (at energies above 2 keV).
In spite of the fact that  no pulsed emission is
detected from the point-like source in either  X-ray or in radio
frequencies, the presence of a neutron star powering the central
emission is implied from the spatial and spectral analysis. 

{\it Plucinsky} summarized observations of {\bf MSH 15-56}. This remnant 
is
similar to MSH 11-62 in the sense that it also belongs to the
``true'' radio-composite class, \ie with a central radio nebula with 
flat
spectrum ($\alpha \sim -0.1$) surrounded by a steeper spectrum shell
($ \alpha \sim -0.4$). {\it ASCA} observations have revealed that in 
addition to
the central  thermal X-ray emission, already known from studies
based on {\it ROSAT} observations, non-thermal radiation is present in a 
small
localized region on the SW shell  partially overlapping but  
slightly offset from the peak of the radio plerion.  {\it Plucinsky} 
suggests
that an unseen moving pulsar may have produced the radio trail, and the
hard X-ray emission indicates the current location of the pulsar.

{\it Long, Blair \& Winkler} presented  {\it ROSAT PSPC} observations of 
four
``X-ray centrally-concentrated'' SNRs: {\bf HB 3}, {\bf
HB 9}, {\bf HB 21} and {\bf W 63}. All of them show
considerable internal structure. The X-ray radiation appears
to  fill all the region within the radio
shell in the cases of HB 3 and HB 9. The emission from HB 21 and W 63 
does
not fill the radio shell, but this may be a consequence of the fact that
these SNRs are fainter, and the brightness of the outer portions may 
fall
below the surface brightness limit. 
The spectra of  all  four remnants are similar, peaking sharply at
about 0.9 keV, most likely arising from thermal X-ray emission from
plasmas with normal abundances. \ha ~ and \sii~ images of the remnants
show a poor correlation with the X-ray images. The thermal energy 
content
of the hot gas detected in HB 3 and HB 9 was found to be close to the
typical values; for HB 21 and W63, the energy content was far less,
however, consistent with the suggestion that these two SNRs are well
into the radiative phase of their evolution.

\subsubsection {The interaction of ``thermal X-ray composites'' with the
environs}

{\it Rho} called attention to the fact that many of the X-ray 
thermal composite remnants are interacting with molecular clouds and
suggested that this may be a property of the class. Particularly, {\it
Rho} showed the case of interaction of the SNRs
{\bf 3C 391} and {\bf W 44} with molecular clouds 
based on observations of the infrared \oi ~ 63
$\mu$m line (Reach \&\ Rho 1996). The infrared lines are reported to be brighter
at the edges of the remnants, suggesting pre-shock densities greater
than $10^3~\rm{cm}^{-3}$. Also continuum IR emission from dust heated by
the shock was detected. The molecular cloud  which the shock front
of 3C 391 is currently impacting was mapped in millimetric spectral 
lines
of CS, HCO$^+$ and $^{12}$CO.

{\it Goss} reported the detection of  OH(1720 MHz) masers associated 
with
SNRs. This is particularly relevant for the study of ``thermal X-ray
 composites'' since for all the members of this class that were searched
for OH(1720 MHz) masers, the result was a positive detection. The sample
includes the SNR {\bf W 28}, where 41 individual OH masers were 
reported;
 {\bf W 44}, where 25  features were detected in six regions spread
out over $30'$, the majority located near the region where a 
dense molecular cloud was  reported to be in contact with
the remnant; {\bf IC 443} for which  6 OH(1720 MHz) features have been 
detected; {\bf 3C 391}; {\bf W 51 C}; etc.

The importance of these searches is that the OH(1720 MHz) masers are
an unmistakable shock signature that can act as
pointers for SNRs in dense molecular environments. This occurs 
because the masers 
are  collisionally excited   by H$_2$ molecules in post
shock gas heated by non-dissociative shocks. 

\subsection { Theoretical Results}\label{sec:composite_theory}

In order to explain SNRs with  interior-peaked X-rays  and a limb-
brightened
 radio shell, White \& Long (1991) proposed the
presence of a mass reservoir in the center to increase the central 
emission.
They developed a similarity solution based on a two-phase structure of
the interstellar medium, with clumps and inter-clump gas. The ISM clumps
engulfed by the SNR would be left relatively intact after the passage of
the shock and slowly evaporate in the interior via thermal conduction.
The similarity solution depends on two dimensionless parameters: {\it 
C},
the ratio of the mean cloud density to the density of the intercloud
material, and $\tau$, the ratio of the cloud evaporation timescale to 
the
remnant age. If the cloud/intercloud density ratio is sufficiently
high and evaporation timescales are relatively long, a morphology with
central brightening in X-ray can be achieved. This solution  reproduces
the observed brightness distribution and temperature profiles for 
several
thermal-composite remnants, assuming  a convenient initial energy for 
the
SN  and adequate density contrast for the surroundings. Sometimes,
however, the model is in conflict when evaporation timescales exceed by
far the estimated ages for the SNRs, \eg, for
MSH 11-61A  ({\it Slane's}).

An alternative scenario was presented by {\it Shelton}. This model
assumes that saturated thermal conduction transports energy outwards 
from
the very hot center. Thermal conduction reduces the central temperature
and the temperature gradient. Since pressure is unaffected, the central
density is increased to compensate, and the density gradient decreases.
In this way the central density can be about $\sim$ 15\% of the ambient
density, and enough mass is left in the center to make it brighter in
X-rays. In addition, the thermally conductive SNR also emits more
thermal X-ray photons from its interior than its non-conductive
counterpart because the interior temperature has been brought down to 
the
range that produces thermal low-energies X-rays. The 
\underbar{conduction} 
model is fundamentally different from
the \underbar {evaporation} model in the sense that it works in a one-
phase ISM. This model was successfully applied by {\it Shelton} to 
explain
W 44 observations.

Other scenarios proposed to explain the observations suggest: fossil
radiation assuming non-equilibrium ionization ({\it Harrus}), reflected
shocks ({\it Rho}), SN ejecta, differential absorption, 
a three-component ISM, or a combination of several of these effects.

\subsection {Synopsis: Are Thermal X-ray Composites a Class?}

The observational results confirm that this pseudo-class of SNRs with a
radio-shell morphology filled with centrally-peaked X-ray radiation, is 
a
very heterogeneous set of objects, with X-ray emitting masses varying in
a broad range in spite of having similar X-ray spectra ({\it Long}
), or, on the contrary,
dramatically different X-ray spectra for SNRs of similar  morphology and
age ({\it Burrows}). The optical emission
seems to be in general poorly correlated with X-ray features in these
remnants. The surrounding ISM appears to be   strongly affecting the
morphology and evolution of these remnants ({\it Rho} and {\it Goss}).

From this workshop, it can be concluded that multi-wavelength high
quality  observations  (radio, infrared, optical and X-ray imagery
and spectra) are necessary for the galactic candidates, in order to make
a reliable classification of these objects, and especially to understand
the underlying physics and refine the theoretical models. The studies of
the individual objects should be accompanied by surveys of the
surrounding interstellar gas.

Also, as pointed out by {\it Shelton}, the theories elaborated for these 
peculiar remnants will need to 
reproduce  the observed parameters, such as emission profiles, 
masses, temperatures, etc., but also to explain what makes these 
remnants 
appear different from the rest, whether they are an intermediate
evolutionary  stage of ``normal'' SNRs or a consequence of environmental 
factors, and why not all SNRs are thermal-composites.

A final minor issue has been the recognition of the necessity of  
a new name for this heterogeneous class,  descriptive and
inclusive of the many aspects observed. Some suggested acronyms are:
 CPTX ({\bf C}enter {\bf P}eaked {\bf T}hermal {\bf X}-rays SNRs); CCXS 
({\bf
C}entrally {\bf C}ondensed {\bf X}-ray {\bf S}NRs);  INXS ({\bf 
IN}terior 
{\bf X}-ray {\bf S}upernova remnants), etc.

\section{Pulsar-Driven SNRs}\label{sec:pulsar}

Although the focus of the meeting was on shell-type SNRs - those whose
evolution is driven in large part by the initial energy of the supernova
event - there were a number of presentations related to SNRs with
active central energy sources. This
offers another physical situation that can help
us understand the evolution of blast waves and their interactions
with the circumstellar medium. 

\subsection{The Outer Shell of the Crab Nebula}\label{sec:crab}

The SN of 1054 AD that gave rise to the Crab nebula and its pulsar
has long been classed as a type II event. However, the kinetic energy
of the SN, inferred from the mass and velocity of the {\it visible}
material is only a few percent of the canonical value of 10$^{51}$
ergs.  Chevalier (1977) suggested that this energy was carried away in
an unseen high-velocity envelope of ejecta. However, despite sensitive
searches in H$\alpha$, X-rays and radio (Fesen \etal 1997, Predehl \&
Schmidt 1995, Frail \etal 1995) no such shell has been seen. During
this meeting a claimed detection of the interaction between the pulsar
wind with the SN ejecta was greeted with great interest. {\it Hester}
presented the case, which was further elaborated upon by {\it
Sankrit}. {\it HST} observations reveal a thin ``skin'' of [OIII] around 
the
outside of the Crab nebula, which they interpret as a cooling region
behind a radiative shock propagating at $\sim$150 km s$^{-1}$ into
material with a density of $\sim$10 cm$^{-3}$. Their shock model can
explain the brightness of the [OIII] and C IV emission, whereas
photoionization models do not. It is the accelerated pulsar wind that
drives the shock into the inner edge of the slower-moving ejecta.
Thus, only a small amount of the ejecta is illuminated by the shock
and much of the missing material (as well as the outer blast wave)
remains unseen. On the theoretical side {\it Jun} simulated the
wind/ejecta interaction region of the Crab nebula in order to gain
some insight into the formation of the observed
filaments. Instabilities along the shock front give rise naturally to
filamentary structure with the morphology and overall physical
properties as discussed by Hester \etal (1996).

\subsection{Do Naked Plerions Exist?}

While the Crab nebula may have joined the ranks of composite SNRs,
there exist a handful of pulsar-powered nebula that show no evidence
of an outer shell. Is the absence of an outer blast wave (and readily
detectable ejecta) saying something about the environment into which
the remnant is expanding, or is it saying something about the physical
conditions of the SN event? One may argue that the failure to find
these shells at radio wavelengths results because the blast wave is
expanding into a low density medium. In most of these cases the
current surface brightness limits on a shell are severe, and would
place the putative shells in the faintest 10\% of cataloged SNRs.
However, our knowledge of relativistic particle acceleration is still
uncertain enough that one cannot infer the ambient density directly
from these limits (Frail \etal 1995).

{\it Wallace} advocates that the blast waves should not be sought in the 
radio continuum but rather by looking for signs of
interaction with the surrounding atomic and molecular gas. One of the
best cases of an SNR without a limb-brightened shell is G74.9+1.2
(CTB87). Neutral hydrogen observations at 21-cm show that G74.9+1.2
lies within an expanding HI bubble (Wallace \etal 1997). The
continuum morphology of the SNR indicates a flattening along the
northwest edge, at the apparent point of contact between the HI bubble
and the SNR, suggesting that the bubble has impeded the expansion of
the SNR in this direction. The absence of a limb-brightened shell
suggests that no fast ejecta envelope accompanied this SN
explosion. Despite this, {\it Kothes} presented some tentative evidence
for an outer shock on the edge of G74.9+1.2.  Bonn 100-m observations
between 3 and 30 GHz were used to separate out two spectral
components; one a flat-spectrum, pulsar-powered component and another
steeper spectrum component indicative of shocks.

{\it Torii \& Tsunemi} presented X-ray observations of Crab-like and 
composite
SNRs made with the {\it ASCA} satellite. For the well-known SNR 3C58 
these
observations can be used to put sensitive limits on any thermal
emission from an outer blast wave. The {\it ASCA} spectrum shows no 
emission
lines and a power law fit to the spectrum constrains the brightness of
any thermal components. Taken together, this puts tight limits on the
amount of thermal plasma swept up by the blast wave or heated by a
reverse shock. Deep H$\alpha$ limits have also been placed on a
putative shell around 3C58 by {\it Fesen} in much the same manner as
earlier work on the Crab nebula (Fesen \etal 1997). 3C58 shows a
number of unusual features compared to the Crab nebula. While both
remnants are of a similar age, 3C58 is approximately twice as large as
the Crab, and yet their expansion velocities, inferred from optical
emission lines, are comparable. One interpretation is that the optical
emission in 3C58 is from circumstellar gas, and the true ejecta from
the SN has yet to be detected.

\subsection{Compact Objects in SNRs}

An increasing amount of information on compact objects in SNRs
results from the success of the {\it ASCA} satellite, with its high 
resolution
spectroscopy and imaging capabilities extending to energies of 10 keV.
One notable example is the compact object 1E 161348$-$5055, first
detected by the Einstein satellite interior to the SNR RCW103. {\it
Gotthelf} (Gotthelf, Petre, \&\ Hwang 1997) showed that beyond 3 keV 
the thermal emission from the SNR
is much reduced and the point source can be easily seen. Its spectrum
is best fit by a black body with a temperature kT=0.6 keV. 1E
161348$-$5055 may be a prototype of an emerging class of radio-quiet
neutron stars found interior to shell-type SNRs but with no signs of a
pulsar-powered nebula. Other possible members of this class include
objects in Puppis A (Petre, Becker, \&\ Winkler 1996), 
Kes 73 (Vasisht \&\ Gotthelf 1997; Gotthelf \&\ Vasisht 1997),
G296.5+10.0 (Vasisht et al. 1997), and G78.2+2.1. None of these
objects has been detected as a radio pulsar, and the absence of bright
pulsar-powered nebulae suggest periods and magnetic fields well
outside the range of birth values for pulsars (Frail 1997). These
intriguing objects may be giving us powerful clues about the zoo of
possible progeny of core-collapse SN. {\it Dubner, Goss, Mirabel \& Holdaway} 
presented new multi-band results on SS433/W50.
They find evidence for precession of SS433's jets, even on large scales,
They are studying
the interactions between the jets and the extended remnant, and the 
remnant and its neutral hydrogen environment.

\section{Particle Acceleration}\label{sec:accel}

The link between SNRs and particle acceleration goes back almost
half a century to the suggestion by Shklovsky (1953)
that optical continuum emission from the Crab nebula might be
synchrotron radiation from relativistic electrons. That remarkable, 
early 
insight proved correct. The origin of relativistic electrons
in pure shell SNRs is probably very different from that
in the Crab, but today we recognize the synchrotron process as the 
paradigm for radio emission in those objects, as well. 
({\it Kassim} presented new data
showing extremely uniform radio spectral indices in the Crab, 
consistent with the pulsar as the dominant source of relativistic 
electrons in that object.)
Recently, there has also been 
evidence found for nonthermal X-rays and $\gamma$-rays from a few 
shell SNRs, including some
presented for the first time at this workshop.
The origin of the energetic electrons responsible for 
nonthermal emission in shell SNRs remains a largely unsettled
issue, so considerable discussion in the workshop
addressed that issue.

Especially since the discovery twenty years ago of the 
``diffusive shock acceleration'' (DSA) process, there has been another 
key 
relationship
recognized between SNRs and particle acceleration physics; namely, the
origin of the ionic component of galactic cosmic-rays (CRs). The
simplest, test-particle, steady-state versions of DSA theory make
robust predictions of power-law momentum distributions of 
particles accelerated by the shock. Those power-laws should lead to
a relativistic energy distribution approximating 
$N(E)\propto E^{-2}$ for
strong adiabatic shocks with a density jump near 4. 
After correction for propagation though the ISM, that form may be
consistent with the observed galactic CR
energy distribution.
This spectrum also is similar to the implied mean energy distribution of
relativistic electrons in shell SNRs, although there is a wide
range of spectra actually seen. 
Thus, SNRs have become almost
universally accepted as the source of galactic CRs (usually
extended to the electron component, as well). 
There remain, however, 
important, unresolved issues in that relationship.
Those themes appeared in a number of presentations at the workshop.

\subsection{Diffusive Shock Acceleration}

{\it V\"olk} provided an overview of the links between CR acceleration 
and SNRs
beginning with the energetic connection. He emphasized that
theoretical studies have shown 10-50 \% of the SN explosion energy 
can be left in CRs through DSA in the SNR blast shock. 
This is sufficient to account for the CR energy replenishment rate of
$10^{42}$ erg $s^{-1}$, required by diffusive escape from the galaxy.
In DSA theory charged particles gain energy by repeated
pitch angle scatterings across the velocity jump of the shock.
The scattering MHD waves can be excited by the high energy particles 
themselves.
{\it V\"olk} emphasized that while 
the test-particle limit predictions of the power-law spectrum
have been seen as a great success for the theory, since it seems to be
consistent with the observed CR spectrum,
efficient DSA that transfers a significant
fraction of the energy flux through a shock to
the diffusive CRs requires a fully non-linear treatment.
Several complications arise. The total density jump through the
shock can exceed the nominal factor of 4 expected in a strong, adiabatic
gas shock. On the other hand, this enhanced jump is not discontinuous
on the scale of CR gyroradii,
but includes a smooth precursor due to ``back-reaction'' from 
CRs diffusing upstream. Then CRs with
different scattering lengths will, on average, encounter different
velocity jumps. As also emphasized by {\it Ellison} in his talk on
nonlinear DSA, this obviates the simple, power-law predictions
found for test-particle DSA theory. The sense of the modification is
a hardening of the energy spectrum towards higher energies; \ie , the
spectrum becomes concave below a cutoff determined by escape or
time constraints.
Ellison mentioned that good comparisons have now been made between
direct measurements of particle spectra in heliospheric shocks and 
predictions 
of 
nonlinear DSA theory using Monte Carlo and diffusion-convection
methods (\eg, Baring \etal 1995; Kang \& Jones 1997).

The key unresolved problems in basic DSA theory itself are an 
understanding
of CR ``injection'' out of the thermal plasma at shocks and the 
heating of the gas and the associated dissipation of MHD waves.
No new results regarding the injection problem were reported
at the workshop, but {\it V\"olk} pointed to
recent theoretical progress on the ``thermal leakage'' of ions
in shocks made by Malkov \& V\"olk (1995, 1996). 
Earlier, numerical simulations based on nonlinear Monte Carlo 
techniques, for example, by
Ellison \etal (1996) had shown that ``thermal leakage'' of
the ions into the CR population 
is a natural part of collisionless shock formation. 
The amount of energy held by CRs at different evolutionary stages of
a SNR depends on the shock speed, adiabatic losses and the rate of
increase of the swept-up matter.
{\it V\"olk} emphasized that although CR acceleration during the early 
free-expansion SNR phase could be significant and influence dynamics,
the bulk of galactic CRs detected at Earth should be accelerated during 
the Sedov phase. 
The CRs are presumably mostly released from within the SNR after the 
radiative cooling phase,
when the shock speed eventually drops to the Alf\'ven
speed of the ISM.

The full, nonlinear DSA problem is difficult to compute, especially as 
applied 
to
SNRs, where shocks are neither steady nor planar. 
One of the greater technical difficulties in full DSA theory for SNRs 
comes from the likelihood that the CR scattering lengths range
over several orders of magnitude from the thickness of the gas 
subshock to much greater scales. That poses a special problem 
for time-dependent simulations, which are clearly necessary for
SNR calculations.
{\it Berezhko} described nonlinear DSA computations that he and 
collaborators have recently carried out addressing this problem
for the blast waves of SNRs.  
Utilizing
coordinate transformations based on diffusion length scales, 
they successfully computed solutions with CRs obeying 
Bohm diffusion, which assumes scattering lengths to be proportional
to particle gyroradii.
They concluded that the likely source spectrum of CRs escaping
from SNRs in the warm or hot phases of the ISM is proportional to 
$E^{-2.1}$ up to the CR {\it knee} energy,
$\sim 10^{14}$eV (Berezhko \etal 1995; Berezhko \etal 1996;
Berezhko 1996). The observed spectrum at
earth in this energy range has a form proportional to $E^{-2.7}$. The 
commonly
accepted interpretation of isotopic composition in galactic CRs 
as modified by propagation through and escape from the ISM leads to
an energy-dependent ISM column density, $x(E)\propto E^{-0.6}$. So the
Berezhko \etal results seem consistent with observations. 
The models reproduce both the observed galactic CR spectrum
and chemical abundances for ions at energies up to $\sim 1000$ TeV.
The injection rates of protons and ions remain key
issues to be resolved, since they are  still specified {\it ad hoc}.
Berezhko \etal 1996 concluded that the observed enrichment of heavy 
nuclei
compared to solar system abundances can be provided only when the
SNR blast expands into the warm phase ISM.
The SNR shock structures found by Berezhko's group become strongly 
modified 
by the CR back reaction, 
yet they never become smooth, due to the geometrical factors 
in an expanding spherical shock.
That result is consistent with earlier findings based on more
restrictive assumptions (\eg Markiewicz \etal 1990; Kang \& Jones 1991).

Representing another view about the likely source spectrum for
galactic CR, {\it Biermann} reported that he and his colleagues
calculated the expected
$\gamma$-ray emission from CR protons and electrons
and compared it with the diffuse $\gamma$-ray spectrum of our Galaxy
and some SNRs observed by EGRET and the Whipple Observatory 
Cherenkov 
telescope. They found that the {\it source} CR spectrum giving an 
acceptable
fit was $N(E) \propto E^{-2.3}-E^{-2.4}$ instead of $E^{-2.1}$
with the maximum cutoff energy of 10-100 GeV.
Considering that the observed CR spectrum is $E^{-2.7}$,
Biermann argued that this implies the transport of CR may depend
on energy as $E^{1/3}$ rather than the more traditional form, $E^{0.6}$.
Since the maximum cutoff energy of the fitted spectrum
accounting for $\gamma$-rays
is far below the {\it knee} energy, there might be a different
CR {\it source} responsible for CRs
up to the {\it knee} energy, but which contributes insignificantly
to the diffuse $\gamma$-ray emission.
Thus he raised a following question:
Is it possible that
the galactic diffuse $\gamma$-ray emission is produced by the
source CRs accelerated by SNRs expanding into tenuous media, but
encountering dense clouds, while the higher energy
galactic CRs are mostly from the SNRs expanding into stellar winds?
These EGRET data seem to fit very well into the framework
in which the wind-type SNRs are the main acceleration sites for
the galactic CRs up to the {\it knee} (Biermann 1993). 

Using existing X-ray observations and DSA theory, 
{\it Reynolds \& Gaisser} also calculated the maximum 
energy of the electrons in seven shell remnants. 
Their estimated $E_{max}$ for most remnants is $\lsim 50$ TeV, except 
for 
Cas A which has $E_{max}\sim 100$ TeV.
They argued, like Biermann, that the
maximum energy reached by the Sedov phase in these remnants 
seem too low to explain the {\it knee} energy in observed CR spectrum. 
 
{\it Ellison} reported on recent reexamination of CR composition data
and its implications for the origins of heavy elements in the CRs.
This is a very important issue, since it bears directly on the
material from which the CRs are extracted, \ie,
either ejecta from the supernova or from the ISM.
Since the mid-1980s it has been apparent, despite strong enrichment of
heavy elements compared to solar abundances, that detailed
CR abundances do not match the predictions for an origin in 
ejecta enhanced through SN nucleosynthesis.
On the other hand, there is a well-known pattern in the CR abundances 
that has been interpreted since the 1970s as an inverse correlation 
between abundance and the first ionization potential (FIP) for that 
atomic 
species. The idea is that a species is only likely to be injected
into the CR population if it is charged. Composition of the solar
wind seems to reflect such a pattern, so this model for galactic CRs
has been widely accepted.
However, Meyer \etal (1997) have recently shown that a better inverse
correlation
exists between CR abundance and {\it volatility} of atomic species.
Volatility crudely correlates with FIP, but not in detail.
This suggested to Ellison, Drury \& Meyer (1997) that heavy elements
in the CR population may be injected as constituents of interstellar
(refractory) grains, thus reviving an old suggestion by Epstein (1980). 
Ellison and his collaborators pointed out that such grains, 
which should act like ions with a very large mass-to-charge ratio,
may be efficiently accelerated to moderate energies by DSA in supernova 
shocks.
Their Monte Carlo simulations of this process, including simple
corrections for grain destruction, seem to produce a good fit to 
the observed CR abundances.

\subsection{Direct Observational Indicators of Particle Acceleration in 
SNRs}

The potential of the DSA mechanism to explain galactic CR ions
below roughly $10^{14}$eV 
is probably why most theoretical studies have focussed on
proton acceleration in mature/old remnants.
However, there is no direct observational evidence yet 
proving that the CR ions are accelerated in or associated with SNRs.
Established direct observations of nonthermal particles in SNRs 
currently 
involve
only electrons,
primarily through radio 
synchrotron emission.
Recently, however, there has been considerable interest in
the potential of
X-ray and $\gamma$-ray observations as windows to {\it in situ}
particle acceleration in remnants, both for the electronic and
baryonic components. Some detections of nonthermal
emission in these bands have offered encouragement to this effort.
Issues associated with higher energy nonthermal photons were
discussed in a number of papers at the workshop.

\subsubsection{Radio Emissions from Electrons Accelerated in SNRs}

The intense radio synchrotron emission from nonthermal electrons 
provides
direct evidence that electrons are accelerated to relativistic
energies in SNRs. Depending on the local magnetic field, the
represented energies are typically $\sim 1$ GeV.
Although it is widely assumed that these electrons are accelerated
by DSA along with ions, that remains an unproven hypothesis.
There are several difficulties in resolving that question.
The ionic CR source energy spectrum is something close to $E^{-2}$,
as discussed above. The resemblance between that and the limiting
test particle DSA slope for strong shocks was one of the early
arguments for this mechanism in SNRs. While one might naively
expect this to predict the same energy spectral slopes for electrons in 
SNRs,
that is not clearly born out by the synchrotron data, nor is it
necessarily the prediction in a nonlinear DSA model applied to SNRs
at all stages of their evolution.
For example, {\it Ellison} and {\it Reynolds} both emphasized that 
electron
energy spectra may be concave if SNR shocks are modified through 
backreaction from CR ions.
Then the comparison between electron and ion spectral slopes may depend
sensitively on the energies involved.
One basic theoretical difficulty in applying DSA to electrons is that it 
is
still unclear how thermal electrons gain momentum above 
the thermal ions so they can be injected into a population
that can respond to DSA (for nonrelativistic particles at
fixed energy the gyroradius scales as the square root of the mass,
confining low energy electrons on scales smaller than the shock
thickness).

{\it D. Green} summarized observations of radio spectral indices in 
SNRs.
Shell-type SNRs exhibit a broad range of spectral
indices, centered roughly on  $\alpha \approx -0.5$.
This can be related to the electron energy index, $q$ ($N(E) \propto E^{-q}$), 
through the
formula, $\alpha = -(q~-~1)/2$, so that the mean electron spectrum
is roughly $\propto E^{-2}$, as one might expect from the
simple DSA theory for strong shocks. But the range
is from $q\approx 1.4$ to $q \approx 2.6$. The larger values may
be accommodated in the linear theory in terms of weaker shocks
with density jumps of $r \sim 3$, using the linear DSA formula, $r = 
(q~+~2)/(q~-~1)$.
The only way to accommodate the flatter synchrotron spectra within
DSA theory is to include nonlinear effects, so that the total shock 
density
jump experienced by electrons is greater than 4.
On the other hand, {\it Green} pointed out that there is an apparent trend 
in synchrotron spectral index with 
remnant age (more explicitly with diameter).
Young shell SNRs have $\alpha<-0.5$, while old shell SNRs tend to 
have flatter spectra.
The three historical shell remnants, for example all have spectral
indices steeper than $\alpha = -0.6$.
{\it Green} argued that it is difficult, however, to determine the index 
to better than 0.1
in most cases due to inconsistencies among instruments, techniques 
and base levels, and uncertainties in flux density scales. 
He noted that some filled-center SNRs show strong spectral breaks at 
high 
radio
frequencies (\eg 3C58; Green \& Scheuer 1992). 
It is not yet clear if those breaks result from previous
synchrotron losses in an environment with strong magnetic fields
or represent an intrinsic change in the 
injected electron spectrum. They are, however, at frequencies
well below that in the Crab nebula.

Potentially one of the best ways to restrict the nature of
electron acceleration in SNRs would be to measure spatial and/or
temporal variations in the spectra. Then one might hope to
relate the observed spectral properties with the local dynamical
situation or its history. In practice, however, that is fraught with
both observational and theoretical difficulties.
In shell-type SNRs, spectral variations are very complex
in both space and time, and difficult to measure with confidence. 
Despite this, convincing measurements are now appearing.
{\it Landecker} described observational methods that can be used
to map spatial variations of $\alpha$ in several SNRs.
The most obvious method is to make identical
high resolution images at two different frequencies 
and then divide them. But, most high resolution maps come from
aperture synthesis telescopes, where an absence of short baselines
can miss flux in large scale structures. So, in large scale objects 
in complex backgrounds, this does not work well.
Similarly, single antenna observations are prone to background problems.
A better method is the so-called ``T-T plot'', which measures the
differential brightness temperatures (\ie gradient) at two frequencies. 
This 
method
has been applied with success by Rudnick and his students
(\eg Anderson and Rudnick 1993).
{\it Landecker, Higgs, Gray, Zhang \& Zheng } described radio observations of 
G78.2+2.1 
including the spectral index maps derived using this method.
They find variations with a rough axial symmetry around the SNR
with values ranging from $\alpha < -0.7$ on the steep end to $\alpha > -0.4$.
{\it Koralesky \& Rudnick}
described observations of spatial spectral index variations in 
Cas A. The existence of these variations seems
well established now. In addition, there seem to be correlations 
in Cas A between dynamical parameters (such as the expansion factor) 
and spectral index. The physical origins of these patterns are not
yet clear, however.

\subsubsection{X-ray \& $\gamma$-ray Emissions from CR Electrons}

The recent report that non-thermal X-ray emission from the remnant SN 
1006
observed by the {\it ASCA} is likely to be synchrotron radiation from CR 
electrons 
of energy up to 100 TeV (Koyama \etal 1995; Reynolds 1996) 
has led to more X-ray observations of SNRs and discussions of
their theoretical interpretations. These results are especially
tantalizing, since the electron energies implied are comparable
to the {\it knee} in the energy spectrum for CR ions, which 
is commonly seen as representing the maximum energy that can be
produced by DSA in supernova remnant blasts. (See the earlier
discussion of that issue.) Above $\sim 1$ GeV electrons and ions
of the same energy should behave in essentially the same manner during 
DSA.
   
As mentioned earlier, {\it Keohane \etal } reported at the workshop that 
hard X-ray emission in 
one part of
the remnant IC 443 may also be synchrotron radiation (Keohane \etal 
1997). They
suggested that this represents enhanced particle acceleration
in shocks encountering a dense molecular cloud.
That same region has been seen as a shock-excited OH maser source,
as discussed by {\it Goss}.
Another possible nonthermal contributor to the X-ray fluxes in SNRs
is bremsstrahlung, a point made by {\it Vink} and 
previously in the context of hard X-rays from Cas~A by Askarov \etal 
1990. 
{\it Mastichiadis \& de Jager} have calculated $\gamma$-ray synchrotron 
radiation from CR electrons
accelerated in SN 1006 using a time-dependent `onion-shell model'. 
Their estimated electron synchrotron emission fits the radio 
and soft to hard X-ray spectrum quite well.
They also found for SN 1006 that the $\gamma$-ray emission due to 
inverse 
Compton 
scattering of the CMBR is dominant over relativistic bremsstrahlung
and could be detected by the proposed Imaging Atmospheric Cherenkov 
telescopes. 

{\it de Jager \& Mastichiadis} (de Jager \& Mastichiadis 1997) 
suggested that
relativistic bremsstrahlung/inverse Compton scatterings by
electrons associated with SNR W44
can explain the observed $\gamma$-ray emission of the EGRET source
2EG J1857+0118.
Considering its flat electron spectrum, they concluded
the acceleration origin for this remnant is injection by the
pulsar, PSR B1853+01, instead of first-order Fermi shock acceleration.
The electron energy cutoff is about 100 GeV, as
deduced from non-detection of the remnant above 250 GeV . 
The predicted $\gamma$-ray contribution from CR protons is 
smaller than from electrons.
The region of interest appears to represent an interaction with
a dense molecular cloud and has been reported as an OH maser source,
as well.

\subsubsection{$\gamma$-rays from CR Ions in SNRs}

The best opportunity for a clear test of DSA theory applied to ions in 
SNRs may come
through $\gamma$-ray observations.
Inelastic collisions between CR protons and thermal nucleons should 
produce $\gamma$-rays 
via  $\pi^o$ decay.
Dorfi (1991), Drury, Aharonian, \& V\"olk (1994), Naito \& Takahara 
(1994) 
and
Berezhko \& V\"olk (1997), among others, have examined this before
and found that
$\gamma$-ray emission peaks at the beginning of Sedov phase and then
slowly decays. The expected
$\gamma$-ray luminosity  may be too low in most SNRs to be detected 
by {\it EGRET} at $E>100$MeV.
On the other hand {\it V\"olk} emphasized
that because
the expected $\gamma$-ray spectrum is very hard, the
{\it Imaging Atmospheric Cherenkov} telescopes sensitive to air showers
above at $E > 100$GeV may be best able to detect these interactions.
Some recent reports have further stimulated these investigations. For
example, Esposito \etal (1996)
reported several {\it EGRET} source detections above 100 MeV coincident 
with 
SNRs, including 
W44 and IC433. On the other hand, none of the five
{\it EGRET} sources was detected by the {\it CYGNUS} extensive air 
shower 
experiment,
which is sensitive to photons with energies $\sim 100$ TeV (Allen \etal 
1995).
The latter observations placed upper limits near the predictions by 
Drury, Aharonian \& V\"olk (1994).
 
{\it Baring, Ellison \& Reynolds} reported calculations of expected 
$\gamma$-rays
from protons, He, and electrons
accelerated at SNRs based on fully nonlinear Monte Carlo simulations
in plane shocks designed to mimic SNR shock properties.
The $\gamma$-ray emission was calculated for $\pi^0$ decay,
bremsstrahlung, and inverse Compton scattering of background radiation
fields. 
Rather than a power-law spectrum expected from the test-particle model, 
these spectra have curvatures due to the non-linear dynamical effects
alluded to before.
The maximum energy that they compute, limited by the remnant age, is 
typically
1-10 TeV per nucleon for young SNRs in the Sedov phase.
Such a low $E_{max}$ provides natural cutoffs for the $\gamma$-ray 
emission in the TeV range, compatible with the existing upper limits
such as those from {\it CYGNUS} or the {\it Whipple Observatory}.
This implies that $\gamma$-ray bright SNRs may not be major sources of
high energy galactic CRs, but there are other SNRs 
responsible for high energy CRs up to the {\it knee} energy.
This suggestion is consistent with the work reported by {\it Biermann} and
by {\it Reynolds \& Gaisser}.

\subsection{Particle Acceleration in SNRs - How far have we come?}

As summarized by {\it Kang}, it is obvious from discussions in this 
workshop 
that the dynamics of SNRs and associated particle acceleration
physics are complex, especially due to rich interactions between
clumpy stellar and circumstellar medium/ISM. There is concrete 
observational evidence that CR
electrons are accelerated in SNRs, and they are,
in general, consistent with the diffusive shock acceleration theory.
However, 
many observational details, \eg, spectral index variations in
time and space for specific remnants, which remain to be explained by
a fully non-linear, self-consistent treatment. 

In order to explain the observed galactic CRs, the acceleration
of nuclear CRs at SNRs is inevitable in terms of the energy budget of
our Galaxy. 
Predictions of DSA theory seem to explain most essential
aspects of galactic CRs and seem convincingly self-consistent.
The CR physicists have recently put in more efforts to predict the 
$\gamma$-ray emission from $\pi^0$ decay that results from
the proton-nucleon interaction.  
These predictions may soon face direct observational tests,
as $\gamma$-ray observations gain a proper sensitivity. 

Recent observations of SNRs in X-ray and $\gamma$-ray 
and their theoretical interpretations seem to indicate 
that there is a selection effect;
we observe mostly SNRs in dense environment.
According to the standard DSA theory, 
the particles can be accelerated up to the {\it knee} energy at SNRs
only  in a hot tenuous ISM.
This suggests that there are SNRs which we don't observe, in a low 
density 
ISM, and they are the main acceleration sites for the bulk of 
high energy galactic CRs.

\section{Magnetic Fields}\label{sec:mag}

Magnetic fields and their structures in SNRs
are important for several reasons.
For one, strongly turbulent magnetic fields in the vicinity
of shocks are a key element for DSA.
In addition, the brightness of synchrotron emission is at least as
dependent on the strength of the magnetic field as it is on the
concentration of relativistic particles. Thus, to disentangle
acceleration issues from magnetic field structure issues, we must
understand the fields.
Finally, SNRs offer an exceptional laboratory for understanding some
basic issues in magnetohydrodynamics, a central astrophysical problem, 
since they provide
environments with highly conducting, highly dynamical fluids. 

{\it Dickel} summarized observations of magnetic fields around SNRs.
Magnetic fields in the main shell typically show cellular patterns so
that the net linear polarization, for example, is typically only 
a few percent. In most young SNRs the mean field direction has a
clear radial orientation.
But old shell remnants 
display a variety of magnetic field orientations, sometimes including
a mixture of radial and tangential field regions.
The field strength can be estimated by the rotation measures of 
background
radio sources and X-ray observations.
The fields in the SNRs seem  within about an order of magnitude
to be in equipartition with the relativistic electron energy.
However, since the relativistic electron component is energetically
minor and its coupling to local MHD behaviors is unclear,
the reasons to expect an equipartition between these two components
are not obvious.

{\it Jun} discussed MHD simulations of magnetic field generation in 
SNRs.
The magnetic field can be amplified
by the Rayleigh-Taylor (R-T) instability
which results primarily from deceleration of stellar ejecta.
The amplification is especially effective when
either the ejecta or the circumstellar medium is clumpy.
R-T instabilities can explain the radial magnetic field in
the main shell of young SNRs, since it produces long fingers pointed
inward.
MHD simulations of R-T instabilities in a uniform medium
show that the magnetic energy is much less than the turbulent energy.
The blast wave interacting with a clumpy medium can result in additional 
increases in the turbulent energy (Jun \& Norman 1996a, 1996b; Jun \etal 
1997).
{\it Jun \& Jones} presented MHD simulations of a young SNR 
interacting with a cloud of size comparable
to the SNR blast wave, demonstrating vividly that
synchrotron emissivity is sensitive to magnetic field as well as the
electron density, so that observations should be interpreted carefully.
{\it Jung, Jun,  Choe \& Jones } showed preliminary results of 2D simulations of a
young SNR expanding into a uniform medium threaded by a weak,
uniform magnetic field. The shell of SN ejecta is R-T unstable, leading
to an amplification of the magnetic field. But, this amplification is
dependent on the orientation of the field with respect to the shell.
Consequently, the field amplification is strongest in the ``equator''
where the external field is tangential to the SNR blast. Simple
models for synchrotron emission in these shells show an equatorial 
asymmetry very comparable to so-called ``barrel-shaped'' SNRs.

\section{Conclusion}

The many issues discussed in this paper make clear that the
Minnesota SNR Workshop was very intense and wide ranging in
scope, despite a focussed set of objectives.
One of its great successes was the
confluence of people from a diverse set of perspectives, working
with a wide range of investigative techniques.  The value of this
diversity is perhaps best captured in some final thoughts, that
we asked participants to share.  Following are their slightly
edited answers to the questions {\it 1)}~ What is one message
or thought you would like to leave the SNR community from this workshop?
 and {\it 2)}~ What is one result you would like to see, or 
observation/calculation/simulation 
done on SNRs in the next five years? 

\subsection{Messages for the SNR Community}
We should know/note/remember that:\\
\nl$\bullet$ The CSM/ISM is the most important factor determining the 
structure of remnants - from the initial density gradient created by the 
progenitor, to the effects of large- and small-scale ISM structures in 
later evolution.
\nl$\bullet$ Nurture makes a huge difference.  We see many (most?) SNRs 
because they light up their surroundings.
\nl$\bullet$ We need to look at both detailed studies of individual SNRs 
and global properties of SNRs, generally.
\nl$\bullet$ Just because each SNR is unique and every one when studied 
in 
detail is very complicated, we should not abandon attempts to use them 
to establish global models of the phenomenon.
\nl$\bullet$ Keep the overall stellar evolution context in mind when 
interpreting
observations or developing theory.
\nl$\bullet$ While it is, of course, important to study individual 
objects, we
should not lose sight of the common physical processes involved.  There
is a lot of concentration on some objects, which are not typical.  It's
interesting to study them, but it would be better to concentrate on the
physical process than worry so much about classification (\eg thermal
composites).
\nl$\bullet$ Understand prototypical young SNRs
 fully, by multi-$\lambda$ studies,  and use these
as templates to predict evolution over next few 1000 years as function 
of
the progenitor type and associated circumstellar environment (rather 
than 
by the traditional
free-expansion/Sedov/radiative scheme).
\nl$\bullet$ SNRs are predominantly sources of
cosmic rays, more than of thermal energy (``hot gas'') and of ``random''
kinetic energy (turbulence) or photons.  This should have profound 
consequences for galactic halo dynamics (\eg, a galactic wind).
\nl$\bullet$ Cosmic ray acceleration should be efficient enough to 
produce
non-linear dynamic effects -- compression ratios greater than 4; curved, 
not
power law, spectra; precursors where upstream gas is heated and slowed.
$\bullet$ All is not well with our theoretical and numerical exploratory 
SNR models,
and a thorough analysis is necessary to determine what is responsible 
for
this inadequacy; whether it can be solved by higher dimensionality or
a re-analysis of the fundamentally important physical processes going
into models.
\nl$\bullet$ X-ray spectral modeling is a rapidly advancing field; it is 
important to
learn everything we can now from existing X-ray spectra with the caveat
that many subtle effects cannot be properly observed.
\nl$\bullet$ Modeling of thermal X-ray emission from SNRs should 
consider
nonthermal emission mechanisms, which can offset both the continuum
and line strength relative to the continuum.
\nl$\bullet$ Good multi-wavelength observations of large diameter fainter 
SNRs are
needed if we are to understand the effect of environment on ``filled
center", SNRs, {\it a.k.a.} ``thermal composites."
\nl$\bullet$ The community needs more global and collaborative work--but 
don't squelch the technical diversity either.
\nl$\bullet$ Talk to each other more!  We especially need more contacts 
between people working in particle acceleration/cosmic ray physics and 
everyone else in the SNR community. Theorists and observationalists need 
more contact and discussion ({\it n.b.: This was a frequent comment in 
these messages, but not repeated here}).
\nl$\bullet$ It's nice to get on first name terms with SNRs.  There's no 
need 
to
``classify" individual SNRs.  If there is something special about
 a ``numbered" SNR -- give it a name.
\nl$\bullet$ Contrary to the opinion of many that we have ``enough" 
examples of
SNRs at various stages of evolution, it is important to search for
undiscovered SNRs, and to image them as correctly as possible at as
many wavelengths as possible.  Each SNR has unique features, and it is
impossible to say that we have yet recognized all the important
phenomena at work in SNR evolution.  The next SNR we observe may
be the one which produces a paradigm shift.
\nl$\bullet$ Do not concentrate your efforts towards ``pathological 
sources", 
but
rather on apparently ``healthy" ones (nearly symmetric, in a nice
environment, without possible superpositions with other objects).  If
studied deeply enough,  they will not come out to be ``boring
sources", but well-defined challenges to models.  "Pathological
sources" will come later on.
\nl$\bullet$ It is time to seriously consider the possibility that 
filled-
center SNRs may not have associated shells -- that filled center SNRs are 
intrinsically different than ``normal" SNRs.
\nl$\bullet$ There are lots of interesting southern SNRs 
out there still to be studied.
\nl$\bullet$ We have just started to understand SNRs and there is a long 
way to go.

\subsection{Future work on SNRs}

What should be done?
\nl$\bullet$ 3-D simulations are needed as both a test of the 2-D results and to 
analyze how
projection effects can alter our interpretation of both structures and
dynamics.
\nl$\bullet$ Full 3-D simulations including magnetic fields and cosmic 
rays are needed.
\nl$\bullet$ Three-dimensional MHD simulations of the interaction of a 
SNR with an asymmetric ISM or wind-bubble are important.
\nl$\bullet$ Calculate/simulate a sample of wind-blown bubble 
SNRs (if all the SNRs go off in bubbles, what are the sample
characteristics -- $N(> D)$, temperatures, energies one would derive?).
\nl$\bullet$ Study SN interaction with a realistic CSM formed by realistic winds 
from
realistic progenitors.  Binaries are probably important and have been
(theoretically) implicated as progenitors for SNe IIL, IIb, Ib/c.
\nl$\bullet$Make good calculations of SNRs inside realistic cavities,
which will generate X-ray spectra and radio predictions.
\nl$\bullet$ Theorists should model emission  spectra in their
multi-D hydrodynamic simulations, so that observers can
understand how sophisticated their own more modest simulations need
to be in order to extract the essential physics from the data.
\nl$\bullet$ We need a better calculation/estimation of the radio 
emission from SNRs. Even if 
we don't have a good theory for computing synchrotron radiation a
push in estimating radio emission, or even determining the relevant
parameters would be most welcome.
\nl$\bullet$  Further modeling/computational simulations of an expanding 
SNR encountering ISM features (\eg clouds, HII regions, other SNRs, 
bubbles/equities) are important.
\nl$\bullet$ Find the outer shock/edge in the Crab--and show that our 
basic model for this type of object is valid.
\nl$\bullet$ If there is no shell around the Crab, does it make sense 
that 
a low-density
environment could produce it? Carry out high spatial resolution studies 
of 
X-ray spectral variations in SNRs with
{\it AXAF}.
\nl$\bullet$ All else being equal, select observing targets to fill in 
gaps in coverage of individual SNRs.
\nl$\bullet$ Obtain narrow band X-ray imaging of continuum (between lines) at 
$E > 10keV$ in
young SNRs (Cas~A...) at angular resolutions comparable with optical,
radio data.
\nl$\bullet$ Survey low background emission regions to find large 
faint objects (radio X-ray).  Is it possible to have a SNR without an
explosion?  (\eg Crab-like, without shells)  Where are all these SNRs
with a small angular diameter?
\nl$\bullet$ Obtain a clear-cut X-ray observation of a shell-type SNR, plus --
nonthermal radio / X-ray spectra.
\nl$\bullet$ Where are the radio sources in the galaxy associated with 
all the 
Type
Ia's of the last 1000 years?  How do we find them?
\nl$\bullet$ Find the ~0.5M{\sun} of Fe in type Ia (SN1006/Tycho).
How much does X-ray synchrotron emission contribute to X-ray spectra of 
shell SNRs? More multi-wavelength work needs to be done.
\nl$\bullet$ We need more observational determinations of radio spectral index 
variations and
interpretations in terms of particle acceleration theory.  New 
diagnostic
tools are needed to go from the observations to the theory.
\nl$\bullet$ Get and model data from 1987A in all possible ways -- this 
SNR 
remains
a unique opportunity.
\nl$\bullet$ We need to understand the environments of future SNRs.  For 
example,
what is the detailed structure around post-main sequence stars?  Are
these cavities complete?  What are the scales of clumps that remain?
\nl$\bullet$ Find self-similar solutions for different kinds of SNRs evolving 
in 
different
environments.  Starting from more realistic conditions.
\nl$\bullet$ We need a calculation demonstrating stability (or 
instability) of
similarity solutions (\ie if not found to be self similar, does a self 
similar
solution evolve to non-self similar if perturbed).  If so, self-similar
models are not useful.
\nl$\bullet$ Identify the nature of central X-ray sources in 
remnants.
\nl$\bullet$ Progress is needed on understanding ``thermal composites."
\nl$\bullet$ Theory/observation:  Understand the variation (within and 
between SNRs) in synchrotron spectra.
Observations:  Measure magnetic field strengths in SNRs (not just
equipartition values).
\nl$\bullet$ ISO-type observations (~1' spatial resolution, high 
spectral 
resolution IR)
of as many SNRs as possible.
\nl$\bullet$ Find a definite answer, from theory, to various microphysical 
processes, like:
1.  Efficiency of cosmic ray injection;\ \
2.  Electrons-ions non-collisional equilibrium; \ \ 
3.  Cosmic ray diffusion and escape probabilities; \ \ 
4.  Efficiency in conduction.\ \ 
We need all possible connections between these points.
\nl$\bullet$ Measure spectral lines of refractory elements upstream and 
downstream of shocks as tracers of dust grain sputtering--this can be 
used to test gas-dust origin of cosmic rays.  Also we need measurements 
of curved radio spectra.
\nl$\bullet$ A crisp, clear, short article is needed on the different 
cosmic ray production scenarios,
where the subfield is going, and what those of us with hydro codes should
do to better include the effects that CR (production and interaction) 
have
on the hydrodynamics and ionization levels of the gas.
\nl$\bullet$ A new model for filled-center SNRs is needed, updating Reynolds and 
Chevalier
in light of developments since that was published,
and allowing, explicitly, for the possibility of truly ``naked" 
synchrotron
nebula.
\nl$\bullet$ More applications of observations to radiative shock 
simulations
(Sedov/radiative transition) need to be made.

\subsection{Final Comments}

A number of key messages came through clearly at the workshop. The 
importance of the circumstellar medium in regulating the structure and 
evolution of SNRs was a recurring theme.  We began to appreciate that 
the evolution of SNRs was not a simple, one-dimensional path -- that 
individual remnants may be in multiple, concurrent stages of evolution, 
and that multiple, different evolutionary paths probably exist. The 
importance of understanding all of the different thermodynamic and 
physical components of SNRs, including the relativistic plasma, was 
emphasized.  In looking to the future, we saw promising new generations 
of both telescopes and theoretical tools.  And we emerged with an 
enhanced appreciation of how progress critically depends on syntheses of 
information and wide-ranging discussions, such as occurred at this 
workshop.

\acknowledgements
The local organizers (R. Benjamin, R. Dohm-Palmer, T. Jones, B. Jun, B. 
Koralesky, F. Miniati, L. Rudnick)
wish to express their gratitude to all of the many individuals
who helped design and conduct this workshop,
and especially to the scientific organizing committee (Roger Chevalier,
John Dickel, Tatania Lozinskaya, Rob Petre and Heinz V\"olk), 
and to the individual session organizers (Roger Chevalier, Steve 
Reynolds, John Dickel, Dick McCray and Dale Frail). We thank Carl Heiles 
for drafting the section on galactic loops.
Financial support for this workshop was provided by the University of
Minnesota Department of Astronomy as part of its centennial celebration
during 1996-97.

The authors recognize the support for their own work, as follows: 
Work on SNRs and Cosmic Ray acceleration at the University of
Minnesota (BJ, TJ, LR) is supported by the NSF, NASA and the
University of Minnesota Supercomputing Institute.
The VLA is a facility of the National Science Foundation operated under
cooperative agreement by Associated Universities, Inc. (DF).
Basic research in radio and infrared astronomy at the 
Naval Research Laboratory is supported by 
the Office of Naval Research (NK). Work on SNRs at the North Carolina State
University (KJB) is supported by NASA and the North Carolina Supercomputing
 Center.

\section{Appendix - Key to Author References in Text}

The following list includes all names and institutions
 of people who contributed to
the meeting.  Those marked with an asterisk were in attendance to
present their individual or collaborative work.

\noindent *Allen, Glenn E.---National Research Council; ---NASA Goddard Space Flight Center\\
\noindent *Anderson, Martha---University of Wisconsin\\
\noindent *Arnett, David---University of Arizona\\
\noindent *Bandiera, Rino---Osservatorio Astrofisico di Arcetri, Italy\\
Baring, Matthew---NASA Goddard Space Flight Center\\
\noindent *Benjamin, Robert---University of Minnesota\\
\noindent *Berezhko, Evgeny---Institute of Cosmophysical Research \& Aeronomy, Russia\\
\noindent *Biermann, Peter ---M.P.I. f\"ur Radioastronomie, Germany\\
Bietenholz, M.F.---York University, Canada\\
Blair, William P.---Johns Hopkins University\\
Bleeker, Johan ---Space Research Organization Netherlands\\
Blondin, John M.---North Carolina State University\\
\noindent *Borkowski, Kazimierz---North Carolina State University\\
\noindent *Burrows, David---Penn State University\\
\noindent *Butenhoff, Chris---University of Minnesota\\
\noindent *Chevalier, Roger---University of Virginia\\
Chi\`eze, Jean-Pierre---DAPNIA, Service d'Astrophysique, Centre d'Etudes de Saclay, France\\
Choe, S.-U.---Seoul National University, Korea\\
\noindent *Christie, Richard, Okanagan University College\\
\noindent *Chu, You-Hua---University of Illinois\\
Cox, D.P.---University of Wisconsin\\
de Jager, O.C.---Potchefstroomse Universiteit vir CHO, South Africa\\
\noindent *Decourchelle, Anne---Service d'Astrophysique, Centre d'Etudes de 
Saclay, France\\
\noindent *Dickel, John, University of Illinois\\
\noindent *Dohm-Palmer, Robbie---University of Minnesota\\
\noindent *Drake, R. Paul---University of Michigan\\
Drury, Luke O'C.---Dublin Institute for Advanced Studies, Ireland\\
\noindent *Dubner, Gloria---Instituto de Astronomia y Fisica del Espacio, Argentina\\
\noindent *Dwarkadas, Vikram V.---University of Washington\\
\noindent *Dyer, Kristy K.---North Carolina State University\\
\noindent *Ellison, Donald---North Carolina State University\\
Ernst Fuerst---M. P. I. f\"ur Radioastronomie, Germany\\
Estabrook, Kent---Lawrence Livermore National Laboratory\\
Ferrara, Andrea---Osservatorio Astrofisico di Arcetri, Italy\\
\noindent *Fesen, Robert---Dartmouth College\\
\noindent *Frail, Dale A.---National Radio Astronomy Observatory\\
\noindent *Gaensler, Bryan---University of Sydney, Australia Telescope National Facility\\
Gaisser, Thomas K.---Bartol Research Institute, University of Delaware\\
Glendinning, S. Gail---Lawrence Livermore National Laboratory\\
Goret, Philippe---DAPNIA, Service d'Astrophysique, Centre d'Etudes de Saclay, France\\
\noindent *Goss, W. Miller---National Radio Astronomy Observatory\\
\noindent *Gotthelf, Eric V.---Universities Space Research Association,
NASA Goddard Space Flight Center\\
Graham, James R.---University of California-Berkeley\\
Gray, A.D.---Dominion Radio Astrophysical Observatory, Canada\\
\noindent *Green, Anne J.---University of Sydney, Australia\\
\noindent *Green, Dave---Mullard Radio Astronomy Observatory, UK\\
Grenier, Isabelle---DAPNIA, Service d'Astrophysique, Centre d'Etudes de Saclay, France\\
Guo, Zhiyu---NASA Goddard Space Flight Center\\
\noindent *Harrus, Ilana---Harvard-Smithsonian Center for Astrophysics\\
\noindent *Heiles, Carl---University of California-Berkeley\\
Heinrichsen, I.---M. P. I. f\"ur Kernphysik, Germany\\
Helfand, D.J.---Columbia University\\
\noindent *Hester, J. Jeff---Arizona State University\\
Higgs, L.A.---Dominion Radio Astrophysical Observatory, Canada\\
Holdaway, Mark---National Radio Astronomy Observatory\\
\noindent *Huang, D.---NASA Goddard Space Flight Center\\
\noindent *Hughes, John P.---Rutgers University\\
Hurford, Alan---Dartmouth College\\
\noindent *Hwang, Una---NASA Goddard Space Flight Center\\
Jahoda, K.---NASA Goddard Space Flight Center\\
\noindent *Jones, Thomas W.---University of Minnesota\\
\noindent *Jun, Byung-Il---University of Minnesota\\
\noindent *Jung, Hyun-Chul---Seoul National University, Korea\\
Kaastra, Jelle---Space Research Organization Netherlands\\
Kalberla, Peter---Radioastronomisches Institut der Universitat Bonn, Germany\\
Kane, J.---University of Arizona\\
\noindent *Kang, Hyesung---University of Washington\\
\noindent *Kassim, Namir E.---Naval Research Laboratory\\
\noindent *Keohane, Jonathan---NASA Goddard Space Flight Center; University of Minnesota\\
Kesteven, M.J.---Australia Telescope National Facility\\
Kinugasa, K.---Osaka University, Japan\\
Koo, Bon-Chul---Seoul National University, Korea\\
\noindent *Klinger, Robert, J., University of Illinois\\
\noindent *Koralesky, Barron---University of Minnesota\\
\noindent *Kothes, Roland---M. P. I. f\"ur Radioastronomie, Germany\\
Koyama, K.---Kyoto University, Japan\\
\noindent *Landecker, Tom---Dominion Radio Astrophysical Observatory, Canada\\
\noindent *Leahy, Dennis---University of Calgary, Canada\\
\noindent *Levenson, Nancy A.---University of California Berkeley\\
Liang, E.---Rice University\\
Lingenfelter, R.E.---University of California-San Diego\\
London, R.---Lawrence Livermore National Laboratory\\
\noindent *Long, Knox S.---Space Telescope Science Institute\\
\noindent *Lozinskaya, Tatiana---Sternberg Astronomical Institute Moscow State University, Russia\\
Lundqvist, Peter---Stockholm Observatory, Sweden\\
Maciejewski, W.---University of Wisconsin\\
Manchester, R.N.---Australia Telescope National Facility\\
\noindent *Mastichiadis, Apostolos---M. P. I. f\"ur Kernphysik, Germany\\
\noindent *McCray, Richard---University of Colorado\\
Meyer, John-Paul---Service d'Astrophysique, Centre d'Etudes De Saclay, France\\
Milne, Douglas K.---Australia National Telescope Facility\\
\noindent *Miniati, Francesco---University of Minnesota\\
Mirabel, Felix---Service d'Astrophysique, Centre d'Etudes de Saclay, France\\
\noindent *Montes, Marcos---Naval Research Laboratory\\
\noindent *Muxlow, T.W.B.---Jodrell Bank, University of Manchester, UK\\
\noindent *Norman, Michael, L.---University of Illinois\\
Ozaki, M.---Kyoto University, Japan\\
Panagia, Nino---Space Telescope Science Institute; ESA\\
Pawl, A.---University of Wisconsin\\
Pedlar, A.---Jodrell Bank, NRAL, University of Manchester, UK\\
Perley, Rick A.---National Radio Astronomy Observatory\\
\noindent *Petre, Robert A.---NASA Goddard Space Flight Center\\
\noindent *Pineault, Serge---Universite Laval, Canada\\
Plewa, T.---M. P. I. f\"ur Astrophysics, Germany\\
\noindent *Plucinsky, Paul P.---Smithsonian Astrophysical Observatory\\
Rasmussen, I.---Danish Space Science Research Institute\\
\noindent *Raymond, John---Center for Astrophysics\\
Reach, W.T.---Orsay, France\\
Reich, W.---M. P. I. f\"ur Radioastronomie, Germany\\
Remington, Bruce---Lawrence Livermore National Laboratory\\
\noindent *Reynolds, Stephen P.---North Carolina State University\\
Reynolds, J.E.---Australia Telescope National Facility\\
\noindent *Rho, Jeonghee---Service d'Astrophysique, Centre d'Etudes de Saclay, France\\
Roberts, Douglas A.---University of Illinois\\
Robinson, Brian T.---Northwestern University\\
Roger, R.S.---Dominion Radio Astrophysical Observatory, Canada\\
Rothschild, R.E.---University of California-San Diego\\
\noindent *Routledge, David---University of Alberta\\
Rubenchik, A.M.---University of California-Davis\\
\noindent *Rudnick, Lawrence---University of Minnesota\\
Russell, S.---Dublin Institute for Advanced Studies, Ireland\\
Ryu, Dongsu---University of Washington; Chungnam National University, Korea\\
\noindent *Sankrit, Ravi---Arizona State University\\
Sauvageot, J.L.---Service d'Astrophysique, Centre d'Etudes de Saclay, France\\
Schnopper, H.---Danish Space Science Research Institute\\
\noindent *Shelton, R.L.---NASA Goddard Space Flight Center\\
\noindent *Shull, Peter---Oklahoma State University\\
Silchenko, O.K.---Sternberg Astronomical Institute, Russia\\
\noindent *Slane, Patrick O.---Harvard-Smithsonian Center for Astrophysics, France\\
\noindent *Smith, R.K.---NASA Goddard Space Flight Center\\
\noindent *Smith, R.Chris---University of Michigan\\
Snowden, S.L.---NASA Goddard Space Flight Center\\
\noindent *Sollerman, Jesper---Stockholm Observatory, Sweden\\
Soto, Alejandro---Dartmouth College\\
Sramek, Richard---National Radio Astronomy Observatory\\
Staveley-Smith, L.---Australia Telescope National Facility\\
\noindent *Strom, Richard---Netherlands Foundation for Research in Astronomy\\
Swerdlyk, C.M.---Dominion Radio Astrophysical Observatory, Canada\\
Taylor, Russ---University of Calgary, Canada\\
\noindent *Torii, Ken'ichi---Osaka University, Japan\\
Tsunemi, H.---Osaka University, Japan\\
\noindent *Tuffs, Richard---M. P. I. f\"ur Kernphysik, Germany\\
Tzioumis, A.K.---Australia Telescope National Facility\\
Van Dyk, Schuyler---University of California-Los Angeles\\
\noindent *Vink, Jacco---Space Research Organization Netherlands\\
\noindent *V\"olk, H.J.---M. P. I. f\"ur Kernphysik, Germany\\
\noindent *Wallace, Brad---Dominion Radio Astrophysical Observatory, Canada\\
Wallace, R.J.---Lawrence Livermore National Laboratory\\
Weiler, Kurt---Naval Research Laboratory\\
Weimer, S.---University of Calgary, Canada\\
\noindent *Williams, Rosa Murphy---University of Illinois\\
Wills, Karen A.---Jodrell Bank, NRAL, University of Manchester, UK\\
Wilner, D.---Harvard Smithsonian Center for Astrophysics\\
Winkler, P. Frank---Middlebury College\\
\noindent *Wright, Eric---North Carolina State University\\
\noindent *Yusef-Zadeh, Farhad---Northwestern University\\
Zhang, X.---Beijing Astronomical Observatory, China\\
Zheng, Y.---Beijing Astronomical Observatory, China\\

\clearpage
\end{document}